\documentclass[a4paper,12pt]{article}
\usepackage{fancyhdr}

\usepackage{amssymb}

\usepackage{color}
\usepackage{graphicx}

\newcommand{\comma}{\quad , \quad}
\newcommand{\fn}[2]{\mathinner{#1(#2)}}
\newcommand{\keV}{\mathinner{\mathrm{keV}}}
\newcommand{\MeV}{\mathinner{\mathrm{MeV}}}
\newcommand{\GeV}{\mathinner{\mathrm{GeV}}}

\begin{document}

\title{Preheating and Affleck-Dine leptogenesis \\ after thermal inflation}

\author{
Gary N. Felder$^1$, Hyunbyuk Kim$^2$, Wan-Il Park$^2$ and Ewan D. Stewart$^2$ \\[2ex]
$^1$ \textit{Department of Physics, Smith College, Northampton, MA 01063, USA} \\
$^2$ \textit{Department of Physics, KAIST, Daejeon 305-701, South Korea} \\
}

\maketitle

\begin{abstract}
Previously, we proposed a model of low energy Affleck-Dine leptogenesis in the context of thermal inflation.
The lepton asymmetry is generated at the end of thermal inflation, which occurs at a relatively low energy scale with the Hubble parameter somewhere in the range $1 \keV \lesssim H \lesssim 1 \MeV$.
Thus Hubble damping will be ineffective in bringing the Affleck-Dine field into the lepton conserving region near the origin, leaving the possibility that the lepton number could be washed out.
Previously, we suggested that preheating could damp the amplitude of the Affleck-Dine field allowing conservation of the lepton number.
In this paper, we demonstrate numerically that preheating does efficiently damp the amplitude of the Affleck-Dine field and that the lepton number is conserved as the result.
In addition to demonstrating a crucial aspect of our model, it also opens the more general possibility of low energy Affleck-Dine baryogenesis.
\end{abstract}

\thispagestyle{fancy}
\rhead{KAIST-TH 2007/03}

\newpage

\section{Introduction}
\label{intro}

The origin of the observed baryon asymmetry is a fundamental question in modern cosmology.
Forty years ago Sakharov showed that three conditions must be satisfied to generate the asymmetry \cite{Sakharov}, and since that time various models have been proposed to explain it.
But we still don't have any idea which, if any, is correct, because baryogenesis usually depends on unknown physics beyond the reach of current experiments and on the unknown early history of the universe beyond the reach of current observations.

A successful cosmological history must produce necessary relics such as the baryon asymmetry, but must also avoid or dilute unwanted cosmological relics such as gravitinos \cite{KhLi} and moduli \cite{CFKRR}, which can destroy the successful predictions of Big Bang nucleosynthesis \cite{KT} or over-dominate the universe depending on the scale of supersymmetry breaking.
Thermal inflation \cite{Stewart,TI} provides a compelling solution to the unwanted relic problem.
It occurs after the usual high energy inflation, at an energy scale between the intermediate and electroweak scales, releasing enough entropy at a low enough energy scale to dilute the unwanted relics.

However, the usual baryogenesis scenarios do not work well in the context of thermal inflation because of the large entropy production at a low energy scale.
Thus a new model of baryogenesis that fits naturally with thermal inflation is desirable \cite{TIB,SKY}.
In a previous paper \cite{DKWE}, we proposed a simple model based on the superpotential
\footnote{We set $c = \hbar = 8\pi G = 1$ throughout this paper.}
\begin{equation}\label{W}
W = \lambda_u Q H_u u + \lambda_d Q H_d d + \lambda_e L H_d e
+ \lambda_\mu \phi^2 H_u H_d + \frac{1}{2} \lambda_\nu \left( L H_u \right)^2 + \frac{1}{4} \lambda_\phi \phi^4
\end{equation}
where $\phi$ is the flaton whose roll out from the origin ends thermal inflation and generates the Minimal Supersymmetric Standard Model (MSSM) $\mu$-term.
The flaton must also have some unsuppressed couplings to the thermal bath, in order to hold it at the origin during the thermal inflation, but the extra fields involved in these couplings acquire intermediate scale masses after the thermal inflation and hence are only weakly constrained, so we do not include them explicitly.
The baryogenesis is an Affleck-Dine (AD) \cite{Affleck_Dine} type of leptogenesis using $L$, $H_u$ and $H_d$, and triggered by the roll out of the flaton.
In Ref.~\cite{DKWE}, we demonstrated the generation of a lepton asymmetry using a simple homogeneous numerical simulation.
In the usual high energy AD scenarios, the amplitude of the AD field would then be damped by the Hubble expansion, bringing the field into the symmetric region of the potential near the origin, ensuring that the generated lepton number was conserved.
However, in our low energy AD scenario, the Hubble damping is negligible.
Instead, we suggested preheating and thermal friction might damp the amplitude of the homogeneous mode, but testing this idea was beyond the scope of our homogeneous numerical simulation.

During preheating, energy is transferred from a homogeneous oscillating mode to inhomogeneous modes,
resulting in reduction of the amplitude of the homogeneous mode.
Well known mechanisms of preheating include parametric resonance \cite{KLS} and tachyonic instability \cite{FBGK}.
Usually, it is regarded as an intermediate process between the end of inflation and full reheating, i.e.\ thermalization.
However, it can be relevant to the evolution of any condensate, in particular AD condensates.
Especially, it may be a crucial effect for low energy AD scenarios, providing an alternative to the usual Hubble damping.

In this paper, we perform three dimensional lattice simulations to numerically investigate the effect of preheating on the amplitude of the AD fields, and hence on the conservation of the lepton number, in our model.
In Section~\ref{model}, we briefly review our model.
In Section~\ref{lepto}, we describe our scenario for the generation and conservation of the lepton number asymmetry.
In Sections~\ref{sim} and~\ref{results}, we describe and present the results of our numerical simulations.
In Section~\ref{dis}, we discuss interesting points, limitations of our current work, and work to be done in future.

While this work was in progress, and after some of our initial results were reported \cite{Wanil}, another work on the same topic appeared \cite{Kawasaki}.
Their simulation was one dimensional, which restricts the number of modes available for preheating, and they chose different initial conditions and parameters\footnote{Their choice of parameters violates the MSSM constraint $\left( m^2_{H_u} + \left| \mu \right|^2 \right) \left( m^2_{H_d} + \left| \mu \right|^2 \right) < \left| B \mu \right|^2$, but this is not important at this stage of simulations of our model.}.
This may explain why their results look somewhat different from ours, though they come to the same conclusion that the model seems to work.
Unfortunately, there are insufficient details of their simulation in their paper for us to reproduce their results.
We hope to clarify these issues after this paper is published.

\section{The Model}
\label{model}

Our model is based on the superpotential of Eq.~(\ref{W}).
The fourth term generates the MSSM $\mu$-term when the flaton $\phi$ settles down to its vacuum expectation value of the order of $10^{10}$ to $10^{12} \GeV$.
The fifth term generates light Majorana masses for the neutrinos, and the sixth term stabilizes $\phi$.
We assume
\begin{equation} \label{lmunuphi}
\lambda_\mu \sim \lambda_\phi \ll \lambda_\nu
\end{equation}
to give the correct values for $\mu$ and the neutrino masses, assuming a flaton mass of the order of the electroweak scale.

Initially, $\phi$ is held at origin and so the $\mu$-term vanishes. Therefore $H_u$ and $H_d$ lack the usual MSSM $\mu$-term contribution to their mass-squareds.
This allows the possibility that some supersymmetric flat directions involving $H_u$ or $H_d$, in particular $LH_u$ or $H_uH_d$, may initially be unstable, although they are stable in our vacuum.
Thus we consider that $LH_u$, $H_uH_d$ and $\phi$ may have nonzero values, and assume that the other fields remain at the origin \footnote{The non-trivial consistency of this assumption was checked in Ref.~\cite{DKWE}.}.
Then, truncating to a single generation for simplicity, gauge fixing and imposing the D-term constraints, we can parameterize those flat directions as
\begin{equation}
H_u = \left(\begin{array}{c} h_u \\ 0 \end{array}\right)
\comma
H_d = \left(\begin{array}{c} 0 \\ h_d \end{array}\right)
\comma
L = \left(\begin{array}{c} 0 \\ l \end{array}\right)
\end{equation}
with the remaining $D$-term constraint
\begin{equation} \label{D}
D = |h_u|^2 - |h_d|^2 - |l|^2 = 0
\end{equation}
and associated gauge constraint
\begin{equation} \label{J}
J_\mathbf{a} = \sum_{\psi \in \left\{h_u,h_d,l\right\}}  \frac{\psi^* Q D_\mathbf{a} \psi - \left( D_\mathbf{a} \psi \right)^* Q \psi}{i}
= 0
\end{equation}
which are not easily integrated out.
The charges are $Q h_u = h_u$, $Q h_d = - h_d$ and $Q l = - l$, and $D_\mathbf{a} = \partial_\mathbf{a} - i Q A_\mathbf{a}$ is the gauge covariant derivative.
The gauge field $A_\mathbf{a}$ is heavy and so not dynamical.

The potential is
\begin{eqnarray}
V & = & V_0 + m_{H_u}^2 |h_u|^2 + m_{H_d}^2 |h_d|^2 + m_L^2 |l|^2 + m_\phi^2 |\phi|^2
\nonumber \\ & & {}
+ \left[ A_\mu \lambda_\mu \phi^2 h_u h_d + \frac{1}{2} A_\nu \lambda_\nu l^2 h_u^2 + \frac{1}{4} A_\phi \lambda_\phi \phi^4 + \textrm{c.c.} \right]
\nonumber \\ \label{V} & & {}
+ \left| \lambda_\mu \phi^2 h_d + \lambda_\nu l^2 h_u \right|^2
+ \left| \lambda_\mu \phi^2 h_u \right|^2
+ \left| \lambda_\nu l h_u^2 \right|^2
+ \left| 2 \lambda_\mu \phi h_u h_d + \lambda_\phi \phi^3 \right|^2
\end{eqnarray}
$V_0$ is adjusted to give zero cosmological constant and all the other supersymmetry breaking parameters are of the order of the electroweak scale.
We denote a typical electroweak scale mass by $m$.

We assume
\begin{equation} \label{mphi}
m^2_\phi < 0
\end{equation}
so that the thermal inflation ends by $\phi$ rolling away.
We assume
\begin{equation} \label{mLHu}
m^2_L + m^2_{H_u} < 0
\end{equation}
so that $LH_u$ can roll away from the origin before the end of thermal inflation, providing the initial condition for our AD leptogenesis.
This nontrivial assumption results in there being a deeper non-MSSM vacuum \cite{Komatsu} so that our MSSM vacuum is meta-stable.
However, as explained in \cite{DKWE}, this is phenomenologically and cosmologically consistent.
For simplicity, we neglect any field dependent renormalization of the supersymmetry breaking parameters.

\subsection{Physical parameters of the potential}
\label{physpara}

In this section we give formulae for some physically important parameters of the potential.
Note that $m^2_\phi < 0$ and $m^2_L + m^2_{H_u} < 0$.

The magnitude of the flaton vacuum expectation value $\phi_0$ is given by
\begin{equation} \label{phivev}
\left| \phi_0 \right|^2
= \frac{\sqrt{ -12 m^2_\phi + \left| A_\phi \right|^2} + \left| A_\phi \right|}{6 \left| \lambda_\phi \right|}
\sim \left( 10^{10} \mbox{ to } 10^{12} \GeV \right)^2
\end{equation}
The potential energy density at the origin is
\begin{equation}
V_0 = \frac{1}{2} \left( - m^2_\phi + \left| \lambda_\phi \phi^2_0 \right|^2 \right) \left| \phi_0 \right|^2
\end{equation}
The flaton mass-squared eigenvalues at $\phi = \phi_0$ are
\begin{eqnarray}
m^2_r & = & 6 \left| \lambda_\phi \phi_0^2 \right|^2 - 2 m_\phi^2
\\
m^2_a & = & 6 \left| \lambda_\phi \phi_0^2 \right|^2 + 2 m_\phi^2
\end{eqnarray}
where the subscripts $r$ and $a$ denote the radial and angular eigenvectors.

When $\phi = h_d = 0$, the magnitudes of $l$ and $h_u$ at the minima of their potential are given by
\begin{equation} \label{lvev}
\left| l_0 \right|^2 = \left| h_{u0} \right|^2
= \frac{\sqrt{ -6 \left(m^2_L+m^2_{H_u}\right) + \left| A_\nu \right|^2} + \left| A_\nu \right|}{6 \left| \lambda_\nu \right|}
\sim \left( 10^7 \mbox{ to } 10^9 \GeV \right)^2
\end{equation}
When $\phi$ settles down to $\phi_0$, the MSSM $\mu$ and $B$ parameters are generated
\begin{eqnarray}
\mu & = & \lambda_\mu \phi^2_0
\\
B & = & A_\mu + \frac{2 \lambda_\phi^* {\phi_0^*}^4}{|\phi_0|^2}
\end{eqnarray}
and the AD mass-squared eigenvalues at the AD origin become
\begin{eqnarray}
m^2_{L H_u} & = & \frac{1}{2} \left| \mu \right|^2 + \frac{1}{2} \left( m^2_{L} + m^2_{H_u} \right)
\\
m^{2\pm}_{H_u H_d} & = & \left| \mu \right|^2 + \frac{1}{2} \left( m^2_{H_u} + m^2_{H_d} \right) \pm \left| B \mu \right|
\end{eqnarray}
The mass-squared along the $L H_u$ flat direction changes its sign at the critical value $\phi_\mathrm{c}$ given by
\begin{equation} \label{phicritical}
\left| \frac{\phi_\mathrm{c}}{\phi_0} \right|^4 = \frac{ - \left( m^2_L + m^2_{H_u} \right) }{ \left| \mu \right|^2 }
\end{equation}
To estimate the potential energy available in the AD sector, we use
\begin{eqnarray} \label{V1}
V_1 \enskip \equiv & V(0,h_{u0},0,l_0) - V_0 & = \enskip
- \left[ - \frac{1}{2} \left( m^2_L + m^2_{H_u} \right) + \left| \lambda_\nu l_0^2 \right|^2 \right] \left| l_0 \right|^2
\\ \label{V2}
V_2 \enskip \equiv & V(\phi_0,h_{u0},0,l_0) & = \enskip
\left| \mu \right|^2 \left| l_0 \right|^2 + V_1
\end{eqnarray}
where $V(\phi,h_u,h_d,l)$ is the full potential.

\section{Leptogenesis}
\label{lepto}

\subsection{Initial conditions}
\label{leptoini}

Initially, all the fields are held at the origin due to the finite temperature effective potential.
We assume $lh_u$ leaves the origin first.
It will do so via a first order phase transition in which rare bubbles expand to Hubble volume size before percolating, since the expansion rate $H_0 \sim m \left| \phi_0 \right|$ is much less than the scale of the phase transition $m$.
As $l h_u$ rolls away, it gives an increasing mass to the fields it couples to.
These then decay and hence are eliminated from the thermal bath, but deposit their energy, including that of their extra mass gained from $l h_u$, back into the thermal bath.
This could lead to two possibilities for $\phi$'s dynamics, depending on $\phi$'s couplings and the detailed dynamics.

Some fields are removed from the thermal bath, and if these fields couple to $\phi$ then $\phi$'s finite temperature effective potential could be reduced, possibly triggering $\phi$'s phase transition.
In this first case, $l h_u$ and $\phi$ roll away at the same time, though with $l h_u$ in the lead if $\left| m^2_\phi \right| < \left| m^2_L + m^2_{H_u} \right| / 2$.

However, the temperature of the thermal bath is raised which would have the opposite effect, tending to prolong the thermal inflation.
In this second case, $l h_u$ would have time to settle to its minimum, preheat and maybe even decay or be diluted by the remaining thermal inflation, before $\phi$ undergoes its own first order phase transition, again with rare bubbles expanding to Hubble volume size before percolating.

In this paper, we consider the second case as it seems simpler.
The initial condition for the leptogenesis will then be $\phi$'s phase transition ending the thermal inflation, with $l h_u$ in some initial state near its minimum.
In this paper, we will only attempt to crudely mimick $\phi$'s first order phase transition, as described in Section~\ref{simini}.
A bound on the initial state of $l h_u$ can be estimated as follows.

After $l h_u$ rolls away, the energy density in the AD sector is initially
\begin{equation}
{\rho_\mathrm{AD}}_0 \sim m^2 \left| l_0 \right|^2
\end{equation}
The AD sector will decay to the thermal bath at a rate
\begin{equation}
\Gamma_\mathrm{AD} \gtrsim \frac{m^3}{\left| l_0 \right|^2}
\end{equation}
and the rate of transfer of energy from the AD sector to the thermal bath will be
\begin{equation}
\frac{1}{H_0} \frac{d\rho_\mathrm{bath}}{dt} + 4 \rho_\mathrm{bath} = \frac{\rho_\mathrm{AD} \Gamma_\mathrm{AD}}{H_0}
\end{equation}
Therefore the energy transferred in the first Hubble time
\begin{equation}
\Delta\rho_\mathrm{bath} \sim \frac{{\rho_\mathrm{AD}}_0 \Gamma_\mathrm{AD}}{H_0} \gtrsim \frac{m^4}{\left| \phi_0 \right|}
\end{equation}
 will be much larger than $m^4$, and so the decay of the AD sector will be able to maintain the thermal inflation.
The thermal inflation will end when $\rho_\mathrm{bath} \sim m^4$, i.e.\ after
\begin{equation}
\rho_\mathrm{AD} \lesssim \frac{m^4 H_0}{\Gamma_\mathrm{AD}} \lesssim \left| \phi_0 \right| {\rho_\mathrm{AD}}_0 \sim \left( 10^2 \sim 10^8 \right) m^4
\end{equation}

Thus, we can expect the initial fluctuations in $l h_u$ to be fairly small and well preheated if not thermalized.
However, the leptogenesis seems not very sensitive to this initial state, see Section~\ref{results}.

\subsection{Generation of the lepton asymmetry}

When the temperature of the thermal bath drops to around the electroweak scale, it is $\phi$'s turn to roll away from the origin, ending thermal inflation.
As $\phi$ rolls out, $h_d$ acquires linear terms in its potential which force it to become non-zero.
When $\phi$ becomes of the order of its vacuum expectation value, $h_u$ acquires an additional contribution to its mass-squared from the term $|\lambda_\mu \phi^2 h_u|^2$, bringing it, and $l$ and $h_d$, back in towards the origin.
At the same time, the phase of $l h_u$ is rotated by the term $\left( \lambda_\mu \phi^2 h_d \right)^* \lambda_\nu l^2 h_u + \textrm{c.c.}$, generating the lepton asymmetry
\begin{equation} \label{nL}
n_L = \frac{l^* \dot{l} - \dot{l}^* l}{i}
\end{equation}

Note that the nonzero $h_d$ is necessary to generate the lepton asymmetry, and is also essential to stabilize the dangerous field directions $Qd$ and $Le$ which would lead one to the deeper non-MSSM vacuum \cite{DKWE}.

\subsection{Conservation of the lepton asymmetry}

In the usual high energy AD scenario, the generated asymmetry is conserved because the Hubble expansion damps the kinetic energy of the AD field, bringing it into the symmetric region of the potential near the origin.
However, in low energy AD scenarios the Hubble damping is negligible and the asymmetry could be washed out by rescattering off the asymmetric part of the potential away from the origin.
Fortunately, the amplitude of the AD fields can also be damped by the transfer of energy to gradient energy and other fields.
In this paper we focus on the former effect, which generally goes by the name of preheating \cite{KLS,FBGK}.

In our model, the gradient energy of the flaton initially grows very rapidly as the flaton rolls away from the origin, due to the tachyonic potential in the radial direction and the divergence of trajectories in the angular direction.
This rapid growth flattens off as the flaton settles down to oscillating about its minimum, with modest and declining growth of gradient energy due to the non-quadratic terms in the potential around the minimum.
The formation and decay of flaton toplogical defects also has an effect.

The AD potential has a strong dependence on the flaton, so inhomogeneities in the flaton field are transferred to the AD fields, leading to rapid preheating of the AD fields.
In addition to this, preheating due to the couplings among the AD fields themselves is likely to be significant, but it is difficult to analyze analytically due to the complicated form of the AD potential.

The energy of the AD fields can also be transferred to other fields.
When an AD field passes near the origin, some fields which couple to it, such as quarks, may become sufficiently light that they can be thermally populated.
As the AD field moves away from the origin, the fields become heavy again, extracting energy from the AD field.
This energy is then transferred to the thermal bath, initially composed mostly of gluons, as the fields decay.
This results in a friction like effect, which we call `thermal friction', on the motion of the AD fields when they pass near the origin.
The efficiency of thermal friction will depend on how close the AD field has to pass to the origin for the quark or other field to become light enough to be thermally populated, and also on how high the temperature is.
How close the AD field has to pass to the origin will depend on the couplings and the temperature of the thermal bath.
The initial temperature of the thermal bath may be relatively small, perhaps electroweak scale,
but the MSSM has a large hierarchy of Yukawa couplings.
This should allow first generation quarks and leptons to be thermally populated when an AD field passes only moderately close to the origin.
As the temperature rises, quarks and leptons of the higher generations will also be able to be populated.

Acting against these effects, the flaton dependence of the AD potential will tend to transfer energy from the flaton to the AD fields.
The initial flaton energy density $\rho_\phi \sim m^2 \left| \phi_0 \right|^2$ is much larger than the initial AD energy density $\rho_\mathrm{AD} \sim m^2 \left| l_0 \right|^2$, so there is potentially a lot of energy to be transferred.
However, one would expect the energy flow to decrease as the amplitude of the flaton decreases due to preheating, energy transfer and eventually Hubble damping.

The efficiency of these effects will be crucial in determining whether our lepton number asymmetry survives.
However, it is unclear how efficient any of these effects is and even which of them is dominant.
So, to start somewhere, in this paper we analyze the preheating effect numerically, which also includes the energy flow from the flaton to the AD fields.

\subsection{Estimation of the final baryon number asymmetry}

Once the amplitude of the AD fields has been damped sufficiently for the lepton violating terms $\frac{1}{2} A_\nu \lambda_\nu l^2 h_u^2$ and $\left( \lambda_\mu \phi^2 h_d \right)^* \lambda_\nu l^2 h_u + \textrm{c.c.}$ to become negligible, $B-L$ will be conserved.
We denote this initial time by $t_\mathrm{i}$.
The AD fields will complete their decay to the thermal bath while the temperature is still above the electroweak scale, and so sphaleron processes will convert our lepton asymmetry to a baryon asymmetry.
Finally, the flaton decay completes, diluting the baryon asymmetry.
We denote this final time by $t_\mathrm{f}$.

The final baryon asymmetry is given by
\begin{eqnarray}
\frac{n_B}{s} & \sim & \frac{\fn{n_{B-L}}{t_\mathrm{f}}}{\fn{s}{t_\mathrm{f}}}
\sim \frac{\fn{n_{B-L}}{t_\mathrm{f}} \fn{T}{t_\mathrm{f}}}{\fn{\rho}{t_\mathrm{f}}}
\sim - \frac{\fn{n_L}{t_\mathrm{i}} \fn{n_{B-L}}{t_\mathrm{f}} \fn{T}{t_\mathrm{f}}}{\fn{n_{B-L}}{t_\mathrm{i}} \fn{\rho}{t_\mathrm{f}}}
\nonumber \\ \label{nB}
& = & - \left( \frac{m_{LH_u} \fn{n_L}{t_\mathrm{i}}}{V_2} \right)
\left( \frac{V_2}{V_0} \right) \left( \frac{\fn{T}{t_\mathrm{f}}}{m_{LH_u}} \right)
\left( \frac{\fn{n_{B-L}}{t_\mathrm{f}}}{\fn{n_{B-L}}{t_\mathrm{i}}} \frac{V_0}{\fn{\rho}{t_\mathrm{f}}} \right)
\end{eqnarray}
where $\rho$ and $T$ are energy density and temperature of the universe.
We use the factor
\begin{equation}
\frac{m_{LH_u} \fn{n_L}{t_\mathrm{i}}}{V_2}
\end{equation}
to give a dimensionless measure of the efficiency of the lepton asymmetry generation and preservation.
It can be interpreted roughly as the ratio of lepton number density to the total AD number density when the AD fields start oscillating.
We expect
\begin{equation}
\frac{V_2}{V_0} \sim 10^{-10} \mbox{ to } 10^{-2}
\end{equation}
and
\begin{equation}
\frac{\fn{T}{t_\mathrm{f}}}{m_{LH_u}} \sim 10^{-3} \mbox{ to } 10^{-1}
\end{equation}
as described in Section~\ref{physpara} and Ref.~\cite{DKWE}.
The last factor in Eq.~(\ref{nB}) depends on the equation of state of the flaton.
If the preheating of the flaton only populates modes with $k \sim m$, then the flaton would have $p/\rho \ll 1$.
The energy flow from the flaton sector to the AD sector could also affect this factor, but, although this flow is likely significant from the AD point of view, we guess it is not so significant from the flaton  point of view.
With these assumptions, one expects
\begin{equation}
\frac{\fn{n_{B-L}}{t_\mathrm{f}}}{\fn{n_{B-L}}{t_\mathrm{i}}} \frac{V_0}{\fn{\rho}{t_\mathrm{f}}} \sim 1
\end{equation}
Thus, we obtain
\begin{equation} \label{nLovernAD}
\frac{n_B}{s} \sim \left( 10^{-13} \textrm{ to } 10^{-3} \right) \frac{m_{LH_u} \fn{n_L}{t_\mathrm{i}}}{V_2}
\end{equation}
compared with the observational result
\begin{equation}
\frac{n_B}{s} \sim 10^{-10}
\end{equation}

\section{Numerical Simulation}
\label{sim}

We performed a three dimensional lattice simulation with periodic boundary conditions using the algorithm described in the Appendix.
The potential was that of Eq.~(\ref{V}) with constraints Eqs.~(\ref{D}) and~(\ref{J}), and canonical gauge invariant kinetic terms.
The constraint Eq.~(\ref{D}) was modified to
\begin{equation}
D = \epsilon^2
\end{equation}
to cut off the singularity at $h_u = 0$.
The physical parameters and fields were rescaled by a typical electroweak scale mass $m$ and a typical flaton or AD field expectation value $M_\mathrm{TI}$ or $M_\mathrm{AD}$, as described in Ref.~\cite{DKWE}.

\subsection{Simulation parameters}

The lattice volume $L^3$, number of lattice points $N^3$, time step $\Delta t$, and $D$-term constraint singularity cutoff $\epsilon$, were taken as
\begin{equation} \label{latticevars}
L = 50 \, m^{-1}
\comma
N = 128
\comma
\Delta t = 4 \times 10^{-3} m^{-1}
\comma
\epsilon = 5 \times 10^{-3} M_\mathrm{AD}
\end{equation}
We tested our results using different values of these numerical parameters.
Limited computing power constrained us to $N \leq 128$.

The $k$-modes allowed by this lattice are $k = \sqrt{k_1^2 + k_2^2 + k_3^2}$ with
\begin{equation}
k_i = \frac{2\pi n_i}{L}
\end{equation}
where $n_i = 0,1,\ldots,N/2$ with $n_i = 0,N/2$ having degeneracy 1 and the rest degeneracy 2.
This spans the range
\begin{equation}
0.13 \, m = \frac{2\pi}{L} \leq k \leq \frac{\sqrt{3}\,\pi N}{L} = 14 \, m
\end{equation}

The physical parameters were taken as follows.
\begin{equation}
m_{\phi}^2 = -0.217 \, m^2
\comma
m_{H_u}^2 = -1.510 \, m^2
\comma
m_{H_d}^2 = 3.533 \, m^2
\comma
m_{L}^2 = 1.223 \, m^2
\end{equation}
The mass-squareds were given with values up to three decimal places to avoid any accidental resonances and also to fit the MSSM constraints described in Ref.~\cite{DKWE}.
\begin{equation}
\left| A_\phi \right| = 1.0 \, m
\comma
\left| A_\mu \right| = 0.5 \, m
\comma
\left| A_\nu \right| = 0.25 \, m
\end{equation}
\begin{equation}
\left| \lambda_\phi \right| = 1.0 \, m M_\mathrm{TI}^{-2}
\comma
\left| \lambda_\mu \right| = 2.0 \, m M_\mathrm{TI}^{-2}
\comma
\left| \lambda_\nu \right| = 0.2 \, m M_\mathrm{AD}^{-2}
\end{equation}
\begin{equation}
M_\mathrm{AD} = 10^{-2} M_\mathrm{TI}
\end{equation}
\begin{equation} \label{phaseredef}
\arg \left( A_\phi \lambda_\phi \right)
= \arg \left( A_\mu \lambda_\mu \right)
= \arg \left( A_\nu \lambda_\nu \right) = 0
\end{equation}
Note that the phases of Eq. (\ref{phaseredef}) are not physical and can be adjusted by field rotations.
The CP phases were taken as
\begin{eqnarray}
\arg \left( - A_\mu^* A_\nu \right) & = & \left\{
\begin{array}{ll}
\displaystyle \pi - \frac{\pi}{20} & \qquad CP+ \\
\displaystyle \pi & \qquad CP0 \\
\displaystyle \pi + \frac{\pi}{20} & \qquad CP-
\end{array} \right.
\\ \nonumber \\
\arg \left( - A_\phi^* A_\mu \right) & = & 0
\end{eqnarray}

For the choice of parameters above, the formulae of Section~\ref{physpara} become
\begin{equation}
\left| \phi_0 \right| = 0.70 \, M_\mathrm{TI}
\comma
V_0 = 0.23 \, m^2 \left| \phi_0 \right|^2
\end{equation}
\begin{equation}
m^2_r = 6.0 \, m^2
\comma
m^2_a = 5.2 \, m^2
\end{equation}
\begin{equation}
\left| l_0 \right| = 1.7 \times 10^{-2} \left| \phi_0 \right|
\comma
\left| \phi_\mathrm{c} \right| = 0.74 \left| \phi_0 \right|
\end{equation}
\begin{equation}
\left| \mu \right| = 0.97 \, m
\comma
\left| B \right| =  1.5 \, m
\end{equation}
\begin{equation}
m^2_{L H_u} = 0.32 \, m^2
\comma
m^{2+}_{H_u H_d} = 3.4 \, m^2
\comma
m^{2-}_{H_u H_d} = 0.54 \, m^2
\end{equation}
\begin{equation}
V_1 = - 2.6 \times 10^{-4} \, V_0
\comma
V_2 = 8.7 \times 10^{-4} \, V_0
\end{equation}

\subsection{Initial conditions}
\label{simini}

The real initial conditions for the leptogenesis includes the realization of the first order phase transitions of $l h_u$ and $\phi$, or at least the first order phase transition of $\phi$ and the state of $l h_u$ at that time, as discussed in Section~\ref{leptoini}.
During the transitions, bubbles of non-zero field value nucleate, expand and percolate.
The typical size of a bubble just before percolation is expected to be slightly smaller than the Hubble radius.
This introduces a scale much larger than the inverse electroweak scale typical of the rest of the dynamics, making it difficult to simulate both the phase transitions and the leptogenesis dynamics simultaneously.
We leave detailed analytic and numerical investigation of these initial conditions to a subsequent paper, and here just use a simple ansatz that covers a range of possibilities, hopefully without missing too much relevant physics.

We do not include a thermal bath as after the fields roll away from the origin, they decouple from the thermal bath, apart from effects like thermal friction which we plan to consider in a subsequent paper.

We set the initial conditions as
\begin{eqnarray}
\phi & = & \Delta\phi + \delta\phi
\\
l & = & l_0 + \delta l
\\
h_d & = & \delta h_d
\end{eqnarray}
with the initial conditions for $h_u$ determined by
\begin{eqnarray}
D & = & \epsilon^2
\\
\dot{D} & = & 0
\\
\arg h_u & = & \pi/4
\\
A_0 & = & 0
\end{eqnarray}
We choose the initial homogeneous displacement of $\phi$ to be either $\Delta\phi=0$ or $\Delta\phi=4m$.
$\Delta\phi=0$ corresponds roughly to a second order phase transition, and provides one extreme.
A finite value of $\Delta\phi$ is meant to correspond roughly to the displaced field inside a bubble.
$D$ is given by Eq.~(\ref{D}) and $A_0 = 0$ leads via Eq.~(\ref{J}) to
\begin{equation}
j_0 =  \frac{h_u^* \dot{h}_u - \dot{h}_u^* h_u}{i}
- \frac{h_d^* \dot{h}_d - \dot{h}_d^* h_d}{i}
- \frac{l^* \dot{l} - \dot{l}^* l}{i}
= 0
\end{equation}

$\delta\phi$, $\delta l$ and $\delta h_d$ are inhomogeneous fluctuations taken to be the combination of vacuum and thermal fluctuations at a temperature $T \sim m$.
Since, as discussed in Section~\ref{leptoini}, in some cases $\delta l$ may be quite large, we also test the effects of an initial spectrum of $\delta l$ fluctuations with a thermal spectrum of temperature $T \sim 10^6 m$ up to the lattice cutoff $k_\mathrm{max} \sim 10 m$, roughly corresponding to a preheated state with $\sqrt{\langle \left| \delta l \right|^2 \rangle} \sim 10^{-2} \left| l_0 \right|$.

\section{Results}
\label{results}

\subsection{Lepton number asymmetry}

Our most important results are Figs.~\ref{d4l} and \ref{d0l} showing the lepton number asymmetry plotted as a function of time.
In Figs.~\ref{d4l} and \ref{d0l}, the red and blue lines show that a lepton number asymmetry is generated, with sign depending on the sign of the CP violating phases.
The green line is the result of a simulation with CP conserving phases used to estimate the noise level due to the random initial conditions and finite size of the lattice.
In the case of $\Delta\phi=4m$, the angle of initial displacement can contribute to the CP violation and hence affect the lepton asymmetry, as shown in Fig.~\ref{d4lphase}.
We average over this initial angle, corresponding to summing over bubbles with different initial flaton phases, to obtain the lepton asymmetry of Fig.~\ref{d4l}.
A common feature of Figs.~\ref{d4l} and \ref{d0l} is that, although some of the initially generated asymmetry is washed out, a significant fraction is conserved to late times.

The initial conditions, $\Delta\phi=0$ and $\Delta\phi=4m$, correspond to a range of initial $\Delta\phi$ so these results suggest that conservation can be achieved for a wide range of initial conditions.
Moreover, Fig.~\ref{d0lT} shows that the lepton asymmetry is not sensitive to the state of $l h_u$ at the time when the flaton starts to roll away.
This suggests that the fluctuations of the AD fields acquired from those of the flaton through their couplings dominate over any initial fluctuations of $l h_u$.

The lepton asymmetry declines slowly at late times.
This is due to the preheating becoming saturated and energy flow from the flaton sector, as will be discussed in Sections~\ref{preheating} and~\ref{energyflow}.

In Figs.~\ref{d0l} and~\ref{d0lT}, the $CP+$ lepton asymmetries are enhanced after a short period of settling down.
It is unclear what this is due to, but it does not seem to be harmful because it disappears leaving a large enough lepton number asymmetry.
The spatial distribution of the lepton number density shown in Fig.~\ref{d0cpplnumsl} shows a large ball type structure, which seems to be responsible for the strange behavior in Fig.~\ref{d0l}.
We suspect it may be a quasi-stable Q-ball \cite{Coleman,KJ0,KJ}, but it is unclear.
We postpone proper analysis of this to future work.

Thus we have demonstrated the central aim of this paper, that preheating can damp the amplitude of the Affleck-Dine fields sufficiently to conserve the generated asymmetry.
We hope to extend this result to more realistic initial conditions in a future paper.

\subsection{Preheating}
\label{preheating}

The dispersions and spectra of the fields give a more detailed picture of the dynamics leading to the conservation of the lepton asymmetry.
The field amplitudes in Figs.~\ref{d4v} and \ref{d0v} (top left of each figure) show the AD fields settling down into the symmetric region of the potential near the origin, which leads to the conservation of the lepton number.
This is a consequence of very efficient preheating as shown in the graphs of the field dispersions (left middle and bottom) and mean squares about vacuum expectation value (top right) in the figures.
While the decreasing behavior around $t=100$ in the graphs of mean square about vacuum expectation value is due to energy flow from homogeneous modes to inhomogeneous modes, the similar behavior in the dispersion graphs is due to energy flow to higher $k$-modes.

As is clear from Figs.~\ref{d4l} and \ref{d0l}, and the graphs in Figs.~\ref{d4v} and \ref{d0v}, the two cases of $\Delta \phi = 4 m$ and $\Delta \phi = 0$ have different initial transient behaviors.
This is due to the difference in the efficiency of the preheating of the flaton.
For $\Delta\phi=4m$, the flaton has an initial homogeneous displacement larger than the fluctuations,
so that it rolls away along a specific direction with relatively small fluctuations,
leading to slower growth of gradient energy.
This results in the flaton oscillating several times before settling down to its minima,
and causes the violent initial behavior of the AD fields due to their flaton-dependent mass-squared's whose sign also depends on the flaton field value.
However, for $\Delta\phi=0$, the flaton rolls away from the origin with initial fluctuations in all directions.
This leads to rapid growth of gradient energy, resulting in the flaton settling down to its minima within a single oscillation.

Note that the large initial flaton dispersion and its drop around $t=100$ in Fig.~\ref{d0v} is due to the formation and annihilation of flaton strings and walls, which can be seen in Fig.~\ref{d0cppphasesl}.
In contrast, Fig.~\ref{d4cppphasesl} shows that the $\Delta\phi=4m$ case is initially dominated by a single domain, as would be expected due to its homogeneous displacement.
Our model has a $\mathbb{Z}_4$ gauge symmetry identifying the four flaton vacua \cite{DKWE}.
However, a discrete gauge symmetry is more difficult to simulate than a global symmetry, so in our simulation we used a $\mathbb{Z}_4$ global symmetry, and relied on the random initial conditions and finite size of the lattice to get the strings and walls to annihilate, in accord with the dynamics expected for the $\mathbb{Z}_4$ gauge symmetry.

After the initial transient stage, the amplitudes of the AD fields start to grow again.
This is because the energy flow from the flaton sector becomes greater than the rate of energy dispersion to higher $k$-modes due to preheating and thermalization.
This is the reason why the lepton asymmetry in Figs.~\ref{d4l} and \ref{d0l} slowly decreases at late times.
The difference in the late time rate of increase of the AD amplitudes in Figs.~\ref{d4v} and \ref{d0v} is due to difference in the rate of energy flow into the AD sector, which is caused by the difference in field amplitudes at the time when preheating is effectively shut down.
So the efficiency of preheating at the initial transient stage determines the late time behavior of the graphs.
This is shown well by the lattice cutoff dependence shown in Fig.~\ref{d0cppvslattice}.

The spectra in Fig.~\ref{kspec} show the
amplification of $k$-modes through preheating, which damps the homogeneous motion of the flaton and AD fields.
In the flaton sector, we initially have tachyonic growth of low $k$-modes as the flaton rolls away from the origin.
After this initial stage of preheating due to instability, the flaton oscillates around its minimum.
At this stage, nonlinear effects populate slightly higher $k$-modes of the flaton.
At late times, the flaton spectrum is cut off at high $k$, which is consistent with theoretical expectations \cite{FBGK}.
The preheating of the flaton sector is transferred to the AD sector through the flaton-dependent potential of the AD fields.
Remarkably, AD $k$-modes higher than the physical cutoff in the flaton sector are also rapidly populated hitting the high $k$ lattice cutoff before $t=100$ with rapid energy flow from lower to higher $k$-modes.
As a result, the AD lattice modes are fully thermalized for most of the simulation.

Once all the lattice modes have been thermalized, preheating is truncated artificially in the simulation, even though it may continue in reality.
This suggests that the reliability of the lattice results will be compromised at late times, as energy that physically should be flowing into the higher $k$ modes is artificially, numerically shunted into the modes that are present in the grid.
This effect should artifically weaken preheating and we believe it accounts for some of the late time decrease in the lepton asymmetry seen in Figs.~\ref{d4l} and \ref{d0l}.

We have performed experiments with coarser lattices ($N=64,32$) and shown that preheating in the AD sector becomes stronger as the UV resolution is increased, confirming our expectation that our limited resolution is artificially damping the effects of preheating. (These experiments also show that the grid resolution is not affecting the flaton sector.)
Our results thus present a lower bound on the effects of preheating.
Thus in reality one should expect more preheating in the AD sector and hence better conserved lepton number than the simulation shows.
We postpone further study of this issue to future work.

\subsection{Energy flow}
\label{energyflow}

Figs.~\ref{d4cppenergy} to \ref{d0cppenergy} show the components of the energy of the flaton and AD fields as a function of time with total energy densities defined as
\begin{eqnarray}
\label{rhophi}
\rho_\phi & \equiv & \left\langle | \dot\phi |^2 + | \nabla \phi |^2 + V_\mathrm{FL} \right\rangle
\\
\rho_\mathrm{AD} & \equiv & \left\langle \left| D_0 \psi \right|^2 + \sum_{i=1}^{3} \left| D_i \psi \right|^2 + V_\mathrm{AD} \right\rangle
\end{eqnarray}
where $\langle \, \dots \, \rangle$ denotes the lattice average and
\begin{eqnarray}
V_\mathrm{FL}(\phi) & \equiv & V_0 + m_\phi^2 |\phi|^2
+ \left[ \frac{1}{4} A_\phi \lambda_\phi \phi^4 + \textrm{c.c.} \right]
+ \left| \lambda_\phi \phi^3 \right|^2
\\
V_\mathrm{AD}(\phi,\psi) & \equiv & V - V_\mathrm{FL}
\end{eqnarray}

As $\phi$ rolls away from the origin, its potential energy $V_0$ is rapidly converted into kinetic and gradient energy.
Even at late times, the flaton's potential energy is slowly being converted into kinetic and gradient energy, due to continued preheating.
The flaton dependence of the AD potential initially generates potential energy for the AD fields, which is then rapidly converted to kinetic and gradient energy.
At late times, the AD lattice modes have thermalized and the energy from the flaton sector flows into kinetic, gradient and potential energies of the AD fields.
The total energy flow from flaton to AD fields is shown in Figs.~\ref{d4cppenergytot} and \ref{d0cppenergytot}.

The qualitative features of Figs.~\ref{d4cppenergytot} and \ref{d0cppenergytot} can be understood from the energy flow rate into the AD sector
\begin{equation}
\dot{\rho}_\mathrm{AD} = \left\langle \frac{\partial V_\mathrm{AD}}{\partial \phi} \dot{\phi} \right\rangle
\end{equation}
which is obtained by use of the equation of motion of $\phi$.
We can separate the energy flow into three stages.
At the first stage, the AD potential is lifted up causing the AD fields to oscillate with large amplitude.
Then, preheating spreads energy into higher $k$-modes, reducing the field amplitudes.
This results in decreasing energy flow rate.
Finally, the AD sector preheating is saturated due to the lattice cutoff, so energy from the flaton sector starts being accumulated in a fixed number of modes.
This causes increasing field amplitude and therefore increasing energy flow rate.
Thus, the late time behavior of the energy flow depends on the efficiency of preheating, which in turn depends on the initial conditions and the lattice cutoff.
The dependence on the initial conditions explains the different rates of decline in Figs.~\ref{d4l} and \ref{d0l}.

We expect a smaller energy flow for a finer lattice, as shown in Fig.~\ref{d0cppvslattice}, but this will be the case only until a physical cutoff of the AD preheating appears.
If the preheating becomes saturated due to a physical cutoff and the energy flow is still significant,
then we may need the help of thermal friction which could provide another channel for energy to flow away.
We plan to study these issues in a future paper.

\section{Conclusion}
\label{dis}

Recently, we proposed a minimal extension of the MSSM which naturally gives rise to thermal inflation followed by AD leptogenesis.
The expansion rate $H$ at the end of thermal inflation is too low for the usual AD Hubble damping mechanism to conserve the generated lepton asymmetry, but we proposed that preheating and thermal friction may perform the same function.
However, those processes are very complicated and not easy to study analytically even in a simple case.
It is even difficult to know which of these processes will dominate since it depends on the details of the dynamics.

In this paper we numerically analyzed the effect of preheating on the conservation of the lepton number asymmetry in our model.
The lepton number asymmetry is generated as the flaton rolls away at the end of thermal inflation.
It will be conserved if the preheating of the AD sector is efficient enough that the homogeneous oscillation amplitudes of the AD fields are quickly reduced, bringing the fields into the symmetric region of the potential near the origin.

The preheating process starts in the flaton sector by the magnification of the initial fluctuations, that include vacuum and thermal contributions, as the flaton rolls away from the origin and oscillates about the minimum of its potential.
Since the flaton potential has negative curvature near the origin and somewhat larger angle-dependent curvature along angular directions far from the origin, very rapid tachyonic preheating is dominant initially.
When the flaton oscillates around its minimum preheating continues due to the anharmonic terms in its potential (weak parametric resonance).

The AD fields in our model have flaton-dependent masses, so their mass squareds vary with time and even change sign.
This violently transfers inhomogeneities in the flaton sector to the AD sector.
The preheating amplifies high $k$-modes in the AD sector much more rapidly than in the flaton sector, so that the AD fields quickly thermalize over the range of momenta present in our lattice.
If we were able to simulate with higher resolution we would expect to see even more efficient preheating.

The flaton sector also feeds energy into the AD sector.
The energy flow rate depends on the amplitudes of the fields and the time derivative of the flaton.
This means that if the flaton and AD fields are efficiently preheated, resulting in small field values, the energy flow rate will be small.
Since the AD sector preheating is very rapid initially, the energy flow in the initial transient stage is not harmful in terms of lepton number conservation.
But the preheating is quickly saturated, causing the energy flowing from the flaton sector to accumulate in a fixed number of modes.
This causes an increase in the oscillation amplitude of the AD fields, which can wash out the asymmetry, as well as increase the energy flow rate.
In this sense, the balance between the preheating in the AD sector and the energy flow from the flaton sector is crucial for the late time evolution of the lepton asymmetry.

There is a large structure in the spatial distribution of the lepton asymmetry, which may be a quasi-stable Q-ball, but it is not harmful since it soon disappeared without much effect on the asymmetry.
The origin of this object is not clear.
Including the renormalisation of the AD masses should make such objects less stable, assuming it is a type of Q-ball.

The outcome of all these effects is, as the simulations show, that the lepton asymmetry in our model is reasonably well conserved at late times due to the efficient preheating, and is not affected much by the initial state of the AD fields.
The lepton asymmetry decreases slowly at late times but the rate is small enough for one to expect that sufficient lepton asymmetry will be preserved for long enough for other physical processes like thermal friction to start working and stop the decrease.

There are several limitations to this work, which may affect the conservation of the lepton asymmetry.
Firstly, all of our results may depend crucially on the initial conditions.
We found that the initial state of the AD fields has little effect on the lepton asymmetry.
The initial state of the flaton, however, affects the preheating of the AD fields and the energy flow, and thus may affect the conservation of the lepton asymmetry.
A proper treatment of this issue will thus require simulating flaton bubbles.
Once this has been done, it becomes possible to determine the regions of parameter space in which our model works.

The limited size and resolution of the lattice also affect the preheating efficiency.
Since the lattice cutoff is reached early in our current simulation, it restricts the preheating artificially.
A larger and finer lattice would provide more available modes for preheating and thermalization such that it would weaken or even stop the decreasing behavior of the lepton asymmetry.

Finally, we may have to consider thermal friction as another source of damping for the homogeneous mode, if it turns out that preheating is not efficient enough to conserve the lepton asymmetry by itself given realistic initial conditions for the flaton.
We postpone all of these issues to future work.

In summary, our simulation revealed that preheating is efficient and rapid enough to conserve the lepton asymmetry in our low energy AD leptogenesis model, replacing the Hubble damping of the usual high energy AD scenario.

\subsection*{Acknowledgements}

EDS thanks Masahide Yamaguchi for collaboration on an earlier related project.
GNF, WIP and EDS thank the Canadian Institute for Theoretical Astrophysics (CITA) for hosting them during the early stages of this project.
HBK, WIP and EDS thank the Astrophysical Research Center for the Structure and Evolution of the Cosmos (ARCSEC) and CITA for providing computing resources.
HBK, WIP and EDS would like to acknowledge the support from the Korea Institute of Science and Technology Information (KISTI) under `The Strategic Supercomputing Support Program'.
The use of the computing system of the Supercomputing Center in KISTI is also greatly appreciated.
GNF was supported by NSF grant PHY-0456631.
HBK, WIP and EDS were supported in part by ARCSEC funded by the Korea Science and Engineering Foundation (KOSEF) and the Korean Ministry of Science, the Korea Research Foundation grants KRF PBRG 2002-070-C00022 and KRF-2005-201-C00006 funded by the Korean Government (MOEHRD), the KOSEF grant R01-2005-000-10404-0, and Brain Korea 21.

\appendix

\section{Appendix}

\subsection{Gauge invariant adaptive leapfrog with constraints}

In this appendix, we describe our algorithm for numerical integration of our dynamics.
It is a gauge invariant adaptive leapfrog type algorithm with constraints.
It is derived from a discrete action, and so is a variational integrator \cite{VI}, which means that it exactly conserves the constraints and charges, and has good long term energy conservation.
The adaptivity is particularly important to help deal with the singular configuration manifold induced by the $D$-term constraint of Eq.~(\ref{D}).

\subsubsection{Lattice gauge invariance}

The remaining gauge field in our model, corresponding to the $D$-term and gauge constraints of Eqs.~(\ref{D}) and~(\ref{J}), is heavy and so not dynamical, but, unlike the other gauge fields, can not be easily integrated out.

A gauge field is a connection relating variables at two different points in space-time, and so should be considered as a link variable living on a link joining two points.
We will follow the usual way of defining covariant derivatives on a lattice in terms of link variables.

Let's denote nodes by $n$ and links by $l$.
A link $l$ connects from the node $l-$ to the node $l+$.
The links $n-$ flow into, and the links $n+$ flow out of, the node $n$.
The length of a link $l$ is $\Delta_l$.
A link $tl$ is in the time direction and a link $xl$ is in the space direction.

\paragraph{Covariant derivative}

\begin{equation}
U_l \equiv e^{i Q A_l \Delta_l}
\end{equation}
\begin{equation} \label{Dl}
(\mathcal{D}\Phi)_l \equiv \frac{U_l^{-\frac{1}{2}} \Phi_{l+} - U_l^\frac{1}{2} \Phi_{l-}}{\Delta_l}
\end{equation}
\begin{equation}
(\mathcal{D}\Phi)_l^\dagger (\mathcal{D}\Phi)_l = \frac{|\Phi_{l+}|^2 - \Phi_{l-}^\dagger U_l^{-1} \Phi_{l+} - \Phi_{l+}^\dagger U_l \Phi_{l-} + |\Phi_{l-}|^2}{\Delta_l^2}
\end{equation}
where $Q$ is the charge operator, $A_l$ is the gauge field on the link $l$, and $\Phi_n$ is the complex scalar field vector at the node $n$.

\paragraph{Gauge transformation}

\begin{eqnarray}
\Phi_n & \rightarrow & e^{iQ\lambda_n} \Phi_n
\\
A_l & \rightarrow & A_{l} + \frac{\lambda_{l+} - \lambda_{l-}}{\Delta_l}
\\
U_l & \rightarrow & e^{iQ(\lambda_{l+}-\lambda_{l-})} U_l
\\
(\mathcal{D}\Phi)_{l} & \rightarrow & e^{iQ(\lambda_{l+}+\lambda_{l-})/2} (\mathcal{D}\Phi)_l
\end{eqnarray}
where $\lambda_n$ is the gauge transformation parameter at the node $n$.

\paragraph{Gauge fixing}

We set
\begin{equation} \label{gaugefix}
A_{tl} = 0
\end{equation}
by taking $\lambda_{tl+} = \lambda_{tl-} - A_{tl} \Delta_{tl}$.

\subsubsection{Discrete action}

\begin{figure}[t]
\centering
\setlength{\unitlength}{1mm}
\begin{picture}(40,30)(-10,-5)
\multiput(0,0)(10,0){3}{\circle*{1}}
\multiput(0,10)(10,0){3}{\circle*{1}}
\multiput(0,20)(10,0){3}{\circle*{1}}
\put(20,5){\circle{1}}
\put(10,12.5){\circle{1}}
\put(10,15){\circle{1}}
\multiput(0.5,0)(10,0){2}{\line(1,0){9}}
\multiput(0.5,10)(10,0){2}{\line(1,0){9}}
\multiput(0.5,20)(10,0){2}{\line(1,0){9}}
\put(0,0.5){\line(0,1){9}}
\put(0,10.5){\line(0,1){9}}
\put(10,0.5){\line(0,1){9}}
\put(10,10.5){\line(0,1){1.5}}
\put(10,13){\line(0,1){1.5}}
\put(10,15.5){\line(0,1){4}}
\put(20,0.5){\line(0,1){4}}
\put(20,5.5){\line(0,1){4}}
\put(20,10.5){\line(0,1){9}}
\put(2.5,-5){$\Delta x$}
\put(-6,3.5){$\Delta t$}
\end{picture}
\caption{\label{lattice}
Our spacetime lattice.
Hollow nodes are added adaptively to the time links.
The kinetic terms are applied on the adaptive time links, the gradient terms on the space links, the potential on the solid nodes, and the $D$-term constraint on all nodes.
}
\end{figure}
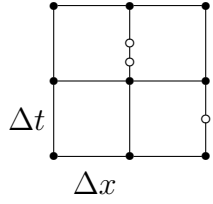

Our discrete action is the sum of kinetic terms on time links, gradient terms on space links, and potential and constraint terms on nodes, with spacetime volume weightings $\Omega$ appropriate to their domains
\begin{equation} \label{action}
S = \sum_{tl} \left( \mathcal{D}\Phi \right)_{tl}^\dagger \left( \mathcal{D}\Phi \right)_{tl} \Omega_{tl}
-  \sum_{xl} \left( \mathcal{D}\Phi \right)_{xl}^\dagger \left( \mathcal{D}\Phi \right)_{xl} \Omega_{xl}
- \sum_{n} V_{n} \Omega^V_{n}
- \sum_{n} \left( D_{n} - \epsilon^2 \right) \Lambda_{n} \Omega^\Lambda_{n}
\end{equation}
where
\begin{equation}
D_{n} = \Phi_{n}^\dagger Q \Phi_{n}
\end{equation}

We choose a regular spacetime lattice with spatial lattice spacing $\Delta x = L / N$ and time step $\Delta t$ for the gradient and potential terms.
This regular lattice is denoted by the solid nodes and the links joining them in Fig.~\ref{lattice}.
Extra nodes, the hollow nodes in Fig.~\ref{lattice}, are added adaptively to the regular lattice's time links, and we use this adaptive lattice for the kinetic and constraint terms.
This helps to follow the dynamics on our constraint manifold which is singular in the limit $\epsilon \to 0$.

For this lattice, the weightings are
\begin{eqnarray}
\Omega_{tl} & = & (\Delta x)^3 \Delta_{tl}
\\
\Omega_{xl} & = & (\Delta x)^3 \Delta t
\\
\Omega^V_{n} & = & \delta_{\bullet n} (\Delta x)^3 \Delta t
\\
\Omega^\Lambda_{n} & \neq & 0
\end{eqnarray}
where
\begin{equation}
\delta_{\bullet n} = \left\{
\begin{array}{ccl}
1 & \textrm{for} & n = \bullet
\\
0 & \textrm{for} & n = \circ
\end{array}
\right.
\end{equation}

\subsubsection{Constraints}

Varying our action of Eq.~(\ref{action}) with respect to the Lagrange multiplier $\Lambda_n$ gives the D-term constraint
\begin{equation}\label{Dcon}
0 = - \frac{1}{\Omega^\Lambda_{n}} \frac{\delta S}{\delta \Lambda_n} = D_{n} - \epsilon^2 = \Phi_n^\dagger Q \Phi_n - \epsilon^2
\end{equation}
Therefore
\begin{eqnarray}
0 = (\partial D)_l & = & \frac{D_{l+} - D_{l-}}{\Delta_l}
= \frac{\Phi_{l+}^\dagger Q \Phi_{l+} - \Phi_{l-}^\dagger Q \Phi_{l-}}{\Delta_l}
\\ \label{dSdL}
& = & \frac{\left(U_l^{-\frac{1}{2}} \Phi_{l+} + U_l^\frac{1}{2} \Phi_{l-}\right)^\dagger Q (\mathcal{D}\Phi)_{l} + (\mathcal{D}\Phi)_l^\dagger Q \left(U_l^{-\frac{1}{2}} \Phi_{l+} + U_l^\frac{1}{2} \Phi_{l-}\right)}{2}
\end{eqnarray}
Varying with respect to the gauge field $A_l$ gives the gauge constraint
\begin{eqnarray} \label{giJ}
0 = - \frac{g_{ll}}{\Omega_{l}} \frac{\delta S}{\delta A_l} = J_{l}
& = & \frac{\Phi_{l-}^\dagger U_l^{-1} Q \Phi_{l+} - \Phi_{l+}^\dagger U_l Q \Phi_{l-}}{i \Delta_l}
\\
& = & \frac{\left(U_l^{-\frac{1}{2}} \Phi_{l+} + U_l^\frac{1}{2} \Phi_{l-}\right)^\dagger Q (\mathcal{D}\Phi)_{l} - (\mathcal{D}\Phi)_l^\dagger Q \left(U_l^{-\frac{1}{2}} \Phi_{l+} + U_l^\frac{1}{2} \Phi_{l-}\right)}{2i}
\nonumber \\ \label{dSdA}
\end{eqnarray}
Combining Eqs.~(\ref{dSdL}) and~(\ref{dSdA}) gives
\begin{eqnarray}
0 = (\partial D + i J)_{l}
& = & \left(U_l^{-\frac{1}{2}} \Phi_{l+} + U_l^\frac{1}{2} \Phi_{l-}\right)^\dagger Q (\mathcal{D}\Phi)_{l}
\\
& = & 2 \Phi_{l+}^\dagger U_l^\frac{1}{2} Q (\mathcal{D}\Phi)_{l}
- (\mathcal{D}\Phi)_{l}^\dagger Q (\mathcal{D}\Phi)_{l} \Delta_l
\\ \label{con}
& = & 2 \Phi_{l-}^\dagger U_l^{-\frac{1}{2}} Q (\mathcal{D}\Phi)_{l}
+ (\mathcal{D}\Phi)_{l}^\dagger Q (\mathcal{D}\Phi)_{l} \Delta_l
\end{eqnarray}

\paragraph{Charge spectrum with \boldmath$Q^3 = Q$}

For $\Phi = (\Phi^+, \Phi^0, \Phi^-)$ with
\begin{equation}
Q \Phi^+ = \Phi^+
\comma
Q \Phi^0 = 0
\comma
Q \Phi^- = - \Phi^-
\end{equation}
Eq.~(\ref{giJ}) simplifies to
\begin{equation} \label{sqequation}
e^{2 i A_{l} \Delta_l} = \frac{\Phi^{+\dagger}_{l-} \Phi^{+}_{l+} + \Phi^{-\dagger}_{l+} \Phi^{-}_{l-}}{\Phi^{+\dagger}_{l+} \Phi^{+}_{l-} + \Phi^{-\dagger}_{l-} \Phi^{-}_{l+}}
\end{equation}
Choosing the positive solution consistent with the limit $\Delta_l \rightarrow 0$, we get
\begin{equation}\label{Asol}
e^{i A_{l} \Delta_l} = \frac{\Phi^{+\dagger}_{l-} \Phi^{+}_{l+} + \Phi^{-\dagger}_{l+} \Phi^{-}_{l-}}{\left| \Phi^{+\dagger}_{l-} \Phi^{+}_{l+} + \Phi^{-\dagger}_{l+} \Phi^{-}_{l-} \right|}
\end{equation}
Note that in our case of $\Phi^+ = h_u$ and $\Phi^- = (h_d,l)$, Eq.~(\ref{Dcon}) gives
\begin{equation}
\left| \Phi^{+\dagger}_{l-} \Phi^{+}_{l+} + \Phi^{-\dagger}_{l+} \Phi^{-}_{l-} \right| \geq \epsilon^2
\end{equation}
We use Eq.~(\ref{Asol}) to determine $U_{xl}$, and from Eq.~(\ref{gaugefix}) we have
\begin{equation} \label{gaugefixU}
U_{tl} = 1
\end{equation}

\subsubsection{Dynamics}

Varying our action of Eq.~(\ref{action}) with respect to $\Phi_{n}^\dagger$ gives
\begin{equation} \label{dSdphi}
0 = - \frac{1}{(\Delta x)^3 \Delta t} \frac{\delta S}{\delta\Phi_{n}^\dagger}
= \widehat{(\mathcal{D}_t^2\Phi)}_n - \delta_{\bullet n} (\mathcal{D}_x^2\Phi)_n
+ \delta_{\bullet n} \frac{\partial V_n}{\partial \Phi_n^\dagger} + \frac{\Omega^\Lambda_{n}}{(\Delta x)^3 \Delta t} \Lambda_{n} \frac{\partial D_{n}}{\partial \Phi_{n}^\dagger}
\end{equation}
where
\begin{equation}
\widehat{(\mathcal{D}_t^2\Phi)}_n
\equiv \frac{U_{tn+}^{-\frac{1}{2}} (\mathcal{D}\Phi)_{tn+} - U_{tn-}^\frac{1}{2} (\mathcal{D}\Phi)_{tn-}}{\Delta t}
\end{equation}
the hat indicates improper normalization due to the use of $\Delta t$ instead of $\Delta_{tn\pm}$, and
\begin{eqnarray}
(\mathcal{D}_x^2\Phi)_n
& = & \frac{U_{xn+}^{-\frac{1}{2}} (\mathcal{D}\Phi)_{xn+} - U_{xn-}^\frac{1}{2} (\mathcal{D}\Phi)_{xn-}}{\Delta x}
\\ \label{cds}
& = & \frac{U_{xn+}^{-1} \Phi_{xn++} - 2 \Phi_n + U_{xn-} \Phi_{xn--}}{(\Delta x)^2}
\end{eqnarray}
Solving for $\Lambda_{n}$ we get
\begin{equation}
\frac{\Omega^\Lambda_{n}}{(\Delta x)^3 \Delta t} \Lambda_{n} n_{n}^\dagger n_{n} =
- n_{n}^\dagger \widehat{(\mathcal{D}_t^2\Phi)}_{n}
+ \delta_{\bullet n} n_{n}^\dagger (\mathcal{D}_x^2\Phi)_{n}
- \delta_{\bullet n} n_{n}^\dagger \frac{\partial V_n}{\partial \Phi_n^\dagger}
\end{equation}
where
\begin{equation}
n_{n} \equiv \frac{\partial D_{n}}{\partial \Phi_{n}^\dagger} = Q \Phi_{n}
\end{equation}
Therefore, defining
\begin{equation}
N_{n} \equiv \left( \frac{n n^\dagger}{n^\dagger n} \right)_{n}
\end{equation}
and
\begin{equation}
P_{n} \equiv 1 - N_{n}
\end{equation}
Eq.~(\ref{dSdphi}) becomes
\begin{equation}\label{Pphi}
P_{n} \widehat{(\mathcal{D}_t^2\Phi)}_{n} - \delta_{\bullet n} P_{n} (\mathcal{D}_x^2\Phi)_{n} + \delta_{\bullet n} P_{n} \frac{\partial V_n}{\partial \Phi_n^\dagger} = 0
\end{equation}

\subsubsection{Algorithm}

Eq.~(\ref{con}) gives
\begin{equation}
N_{n} U_{tn+}^{-\frac{1}{2}} (\mathcal{D}\Phi)_{tn+}
= - \frac{1}{2} \left[ (\mathcal{D}\Phi)^\dagger Q (\mathcal{D}\Phi) \right]_{tn+} \left( \frac{n}{n^\dagger n} \right)_{n} \Delta_{tn+}
\end{equation}
and Eq.~(\ref{Pphi}) gives
\begin{equation}
P_{n} U_{tn+}^{-\frac{1}{2}} (\mathcal{D}\Phi)_{tn+}
= P_{n} U_{tn-}^\frac{1}{2} (\mathcal{D}\Phi)_{tn-}
+ \delta_{\bullet n} P_{n} (\mathcal{D}_x^2\Phi)_{n} \, \Delta t
- \delta_{\bullet n} P_{n} \frac{\partial V_n}{\partial \Phi_n^\dagger} \, \Delta t
\end{equation}
Combining these gives
\begin{eqnarray}
U_{tn+}^{-\frac{1}{2}} (\mathcal{D}\Phi)_{tn+}
& = & P_{n} \left[ U_{tn-}^{\frac{1}{2}} (\mathcal{D}\Phi)_{tn-}
+ \delta_{\bullet n} (\mathcal{D}_x^2\Phi)_{n} \, \Delta t
- \delta_{\bullet n} \frac{\partial V_n}{\partial \Phi_n^\dagger} \, \Delta t \right]
\nonumber \\ \label{pstep0} && {}
- \frac{1}{2} \left[ (\mathcal{D}\Phi)^\dagger Q (\mathcal{D}\Phi) \right]_{tn+} \left( \frac{n}{n^\dagger n} \right)_{n} \Delta_{tn+}
\end{eqnarray}
Implementing our gauge choice of Eqs.~(\ref{gaugefix}) and~(\ref{gaugefixU}) and defining
\begin{equation}
\dot\Phi_{tl} \equiv (\mathcal{D}\Phi)_{tl}
\end{equation}
\begin{equation}\label{pi}
\pi_{n} \equiv P_{n} \left[ \dot\Phi_{tn-} + \delta_{\bullet n} (\mathcal{D}_x^2\Phi)_{n} \, \Delta t - \delta_{\bullet n} \frac{\partial V_n}{\partial \Phi_n^\dagger} \, \Delta t \right]
\end{equation}
\begin{equation}
\Pi_{tl} \equiv \left( \dot\Phi^\dagger Q \dot\Phi \right)_{tl}
\end{equation}
and
\begin{equation}
\tilde\Delta_{n} \equiv \frac{\Delta_{tn+}}{\left( n^\dagger n \right)_{n}}
\end{equation}
reduces Eq.~(\ref{pstep0}) to
\begin{equation} \label{pstep}
\dot\Phi_{tn+} = \pi_{n} - \frac{1}{2} \Pi_{tn+} n_{n} \tilde\Delta_{n}
\end{equation}
while Eq.~(\ref{Dl}) reduces to
\begin{equation} \label{xstep}
\Phi_{tn++} = \Phi_{n} + \dot\Phi_{tn+} \Delta_{tn+}
\end{equation}

Eqs.~(\ref{pstep}) and~(\ref{xstep}) are our basic algorithm, but we still need to determine $\Pi_{tn+}$.
Contracting Eq.~(\ref{pstep}) around $Q$ gives
\begin{equation} \label{it}
\Pi_{tn+} = \left( \pi^\dagger Q \pi \right)_{n}
- \frac{1}{2} \Pi_{tn+} \left[ \left( n^\dagger Q \pi \right)_{n} + \textrm{c.c.} \right] \tilde\Delta_{n}
+ \frac{1}{4} \Pi_{tn+}^2 \left( n^\dagger Q n \right)_n \tilde\Delta_{n}^2
\end{equation}
Choosing the solution which is well behaved in the limit $\tilde\Delta_{n} \to 0$, we get
\begin{equation}\label{Pisol}
\Pi_{tn+}
= \frac{2 \pi_{n}^\dagger Q \pi_{n}}
{1 + \frac{1}{2} \left( n^\dagger_{n} Q \pi_{n} + \textrm{c.c.} \right) \tilde\Delta_{n}
+ \sqrt{\left[ 1 + \frac{1}{2} \left( n^\dagger_{n} Q \pi_{n} + \textrm{c.c.} \right) \tilde\Delta_{n} \right]^2
- \pi_{n}^\dagger Q \pi_{n} n_n^\dagger Q n_n \tilde\Delta_{n}^2} \,}
\end{equation}

For Eq.~(\ref{Pisol}) to remain well behaved we require
\begin{equation} \label{convcon1}
\frac{1}{2} \left| n^\dagger_{n} Q \pi_{n} + \textrm{c.c.} \right| \tilde\Delta_{n} \ll 1
\end{equation}
and
\begin{equation} \label{convcon2}
\left| \pi_{n}^\dagger Q \pi_{n} n_n^\dagger Q n_n \right| \tilde\Delta_{n}^2 \ll 1
\end{equation}
which are the conditions we use for the adaptive part of the algorithm.

For a charge spectrum $Q^3 = Q$, Eq.~(\ref{Dcon}) gives
\begin{equation}
n^\dagger n \geq n^\dagger Q n = \epsilon^2
\end{equation}
Therefore a sufficient condition for Eqs.~(\ref{convcon1}) and~(\ref{convcon2}) is
\begin{equation}
\left| \pi_{n} \right| \Delta_{tn+} \ll \epsilon
\end{equation}
but it is not necessary as typically $n^\dagger n \gg \epsilon^2$.

\subsubsection{Lepton number}

The lepton current is
\begin{eqnarray}
J^L_{l} & = & \frac{\Phi_{l+}^\dagger U_{l}^{\frac{1}{2}} L (\mathcal{D}\Phi)_{l} - (\mathcal{D}\Phi)_{l}^\dagger U_{l}^{-\frac{1}{2}} L \Phi_{l+}}{i}
\\
& = & \frac{\Phi_{l-}^\dagger U_{l}^{-\frac{1}{2}} L (\mathcal{D}\Phi)_{l} - (\mathcal{D}\Phi)_{l}^\dagger U_{l}^{\frac{1}{2}} L \Phi_{l-}}{i}
\\
& = & \frac{\Phi_{l-}^\dagger U_{l}^{-1} L \Phi_{l+} - \Phi_{l+}^\dagger U_{l} L \Phi_{l-}}{i \Delta_l}
\end{eqnarray}
where $L h_u = 0$, $L h_d = 0$ and $L l = l$.
For our gauge choice of Eqs.~(\ref{gaugefix}) and~(\ref{gaugefixU}) the lepton density reduces to
\begin{eqnarray}
J^L_{tl} & = & \frac{\Phi_{tl+}^\dagger L \dot\Phi_{tl} - \dot\Phi_{tl}^\dagger L \Phi_{tl+}}{i}
\\
& = & \frac{\Phi_{tl-}^\dagger L \dot\Phi_{tl} - \dot\Phi_{tl}^\dagger L \Phi_{tl-}}{i}
\\
& = & \frac{\Phi_{tl-}^\dagger L \Phi_{tl+} - \Phi_{tl+}^\dagger L \Phi_{tl-}}{i \Delta_{tl}}
\end{eqnarray}
Our algorithm gives
\begin{eqnarray}
\frac{J^L_{tn+} - J^L_{tn-}}{\Delta t} = \left( \frac{J^L_{xn+} - J^L_{xn-}}{\Delta x} \right) \delta_{\bullet n}
+ i \left[ \Phi_{n}^\dagger L \frac{\partial V_n}{\partial \Phi_n^\dagger} - \frac{\partial V_n}{\partial \Phi_n} L \Phi_{n} \right] \delta_{\bullet n}
\end{eqnarray}
Therefore our algorithm exactly conserves lepton number apart from the lepton number violating terms in the potential.

\subsubsection{Summary}

Our algorithm of Eqs.~(\ref{pstep}) and~(\ref{xstep}), with $\pi_{n}$ given by Eq.~(\ref{pi}), $(\mathcal{D}_x^2\Phi)_{n}$ given by Eq.~(\ref{cds}), $U_{xn\pm}$ determined by Eq.~(\ref{Asol}) and $\Pi_{tn+}$ given by Eq.~(\ref{Pisol}), exactly conserves the $D$-term constraint
\begin{equation}
D_{tl+} = D_{tl-}
\end{equation}
the gauge constraint
\begin{equation}
J_{l} = 0
\end{equation}
and lepton number apart from the lepton number violating terms in the potential, and has good long term energy conservation, as is expected for a variational integrator \cite{VI}.

It is also adaptive with respect to the kinetic and constraint terms, allowing efficient integration on our constraint manifold, which is singular in the limit that the cutoff $\epsilon \to 0$.


\clearpage


\begin{figure}
\includegraphics[width=\textwidth,bb=95 5 320 140]{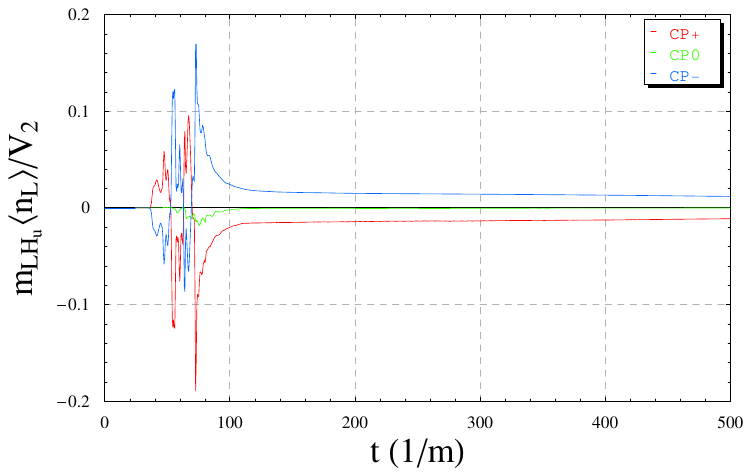}
\caption{\label{d4l}
Lepton number density, averaged over the lattice and initial phase of the flaton, as a function of time, for $\Delta\phi=4m$.
}


\includegraphics[width=\textwidth,bb=100 5 320 145]{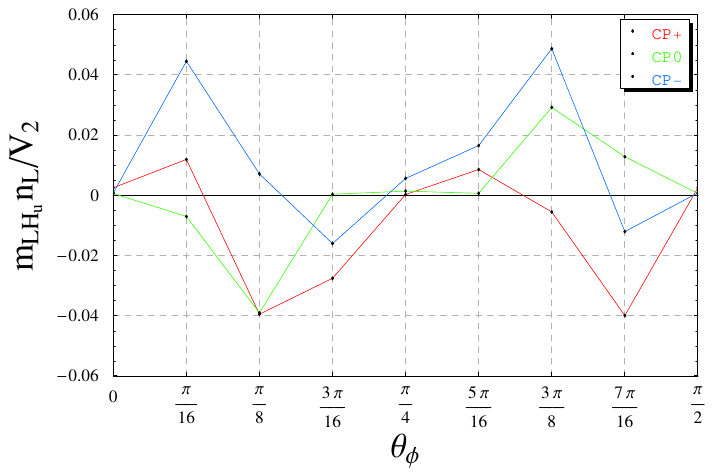}
\caption{\label{d4lphase}
Lepton number density, averaged over the lattice, as a function of the initial phase of the flaton, for $\Delta\phi=4m$.
The $\mathbb{Z}_4$ symmetry of our model means that the range of initial angles is $[0,\pi/2)$.
}
\end{figure}

\clearpage

\begin{figure}
\includegraphics[width=\textwidth,bb=95 5 320 143]{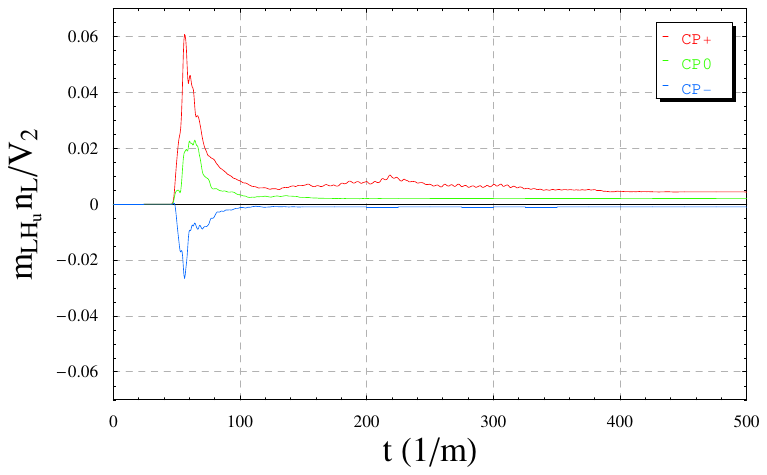}
\caption{\label{d0l}
Lepton number density, averaged over the lattice, as a function of time, for $\Delta\phi=0$.
}



\includegraphics[width=\textwidth,bb=95 5 320 143]{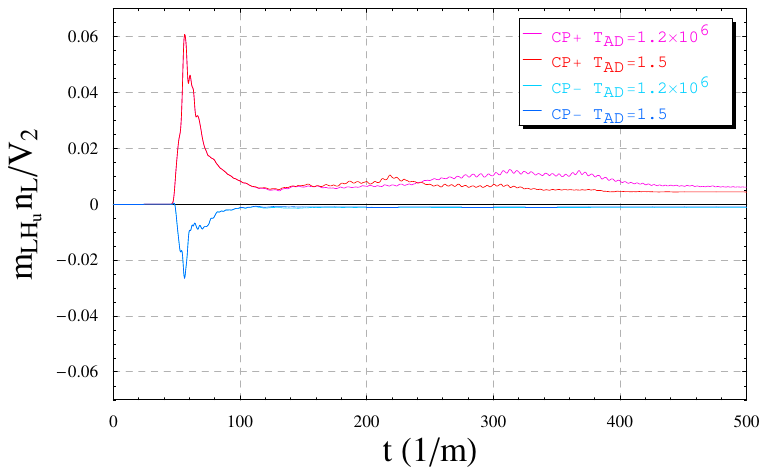}
\caption{\label{d0lT}
Lepton number density, averaged over the lattice, as a function of time, for $\Delta\phi=0$, $CP+$ and $CP-$, with different initial effective temperatures for $l h_u$.
We set an initial thermal spectrum up to the lattice cutoff $k_\mathrm{max} \sim 10 m$, so that an effective temperature of $T \sim 10^6 m$ would correspond to a preheated state with fluctuation $\sqrt{\langle \left| \delta\psi \right|^2 \rangle} \sim 10^{-2} \left| l_0 \right|$.
}
\end{figure}

\clearpage


\begin{figure}
\includegraphics[width=0.5\textwidth,bb=95 3 300 185]{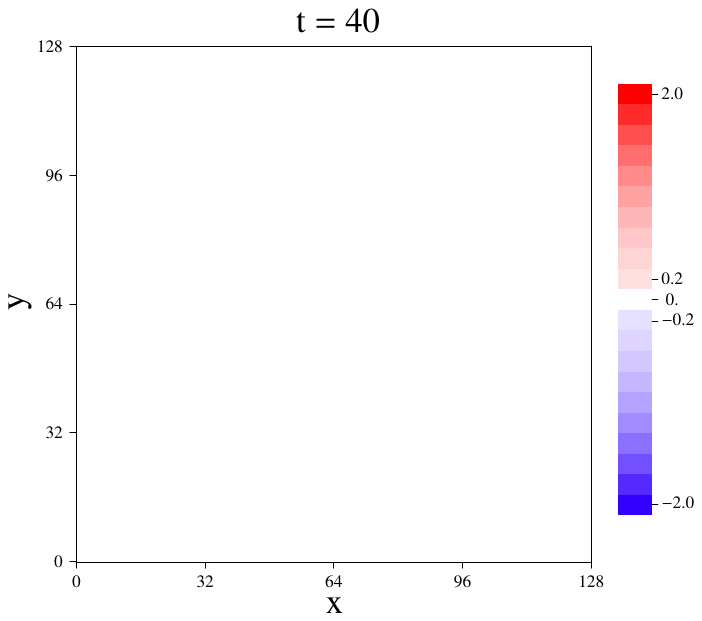}
\includegraphics[width=0.5\textwidth,bb=95 3 300 185]{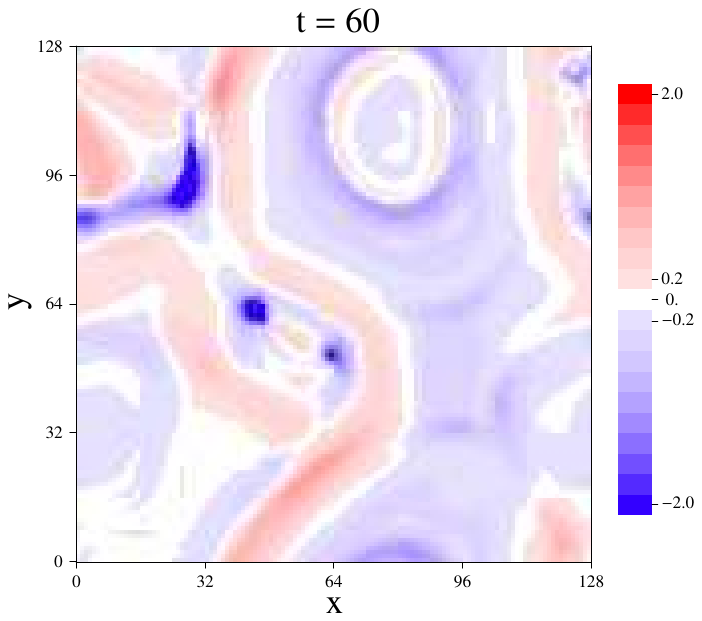}
\includegraphics[width=0.5\textwidth,bb=95 3 300 185]{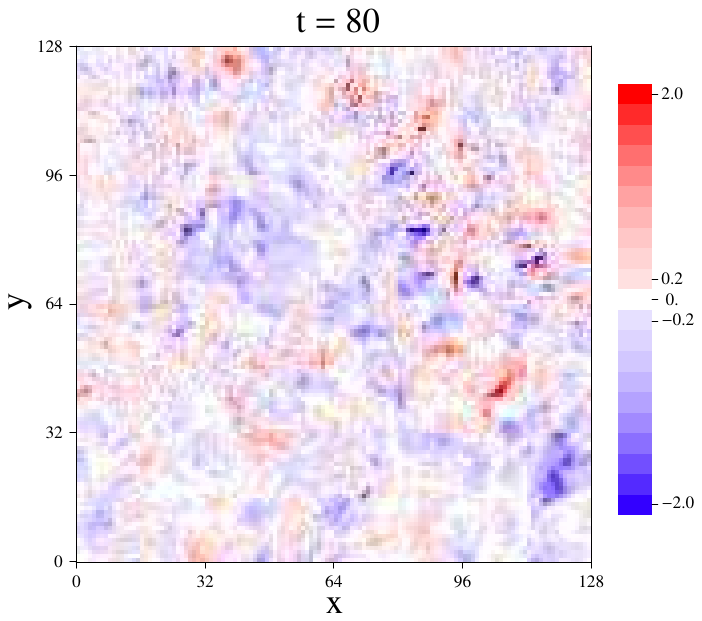}
\includegraphics[width=0.5\textwidth,bb=95 3 300 185]{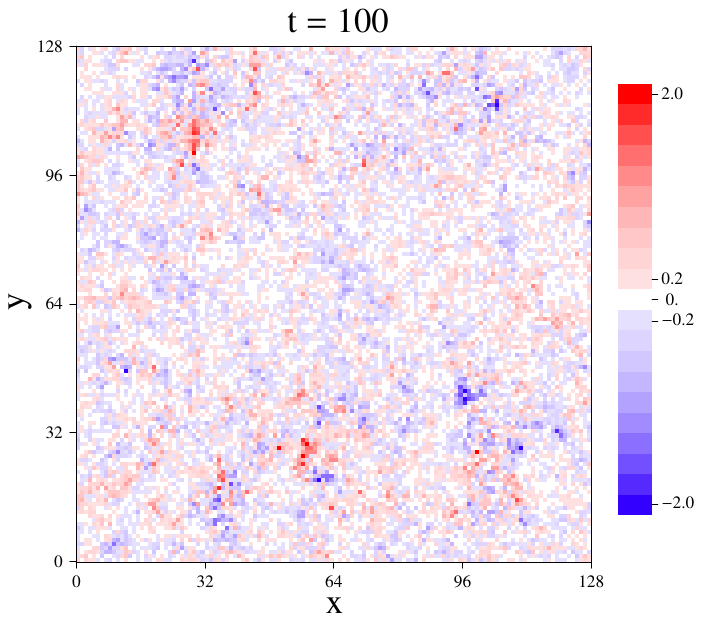}
\includegraphics[width=0.5\textwidth,bb=95 3 300 185]{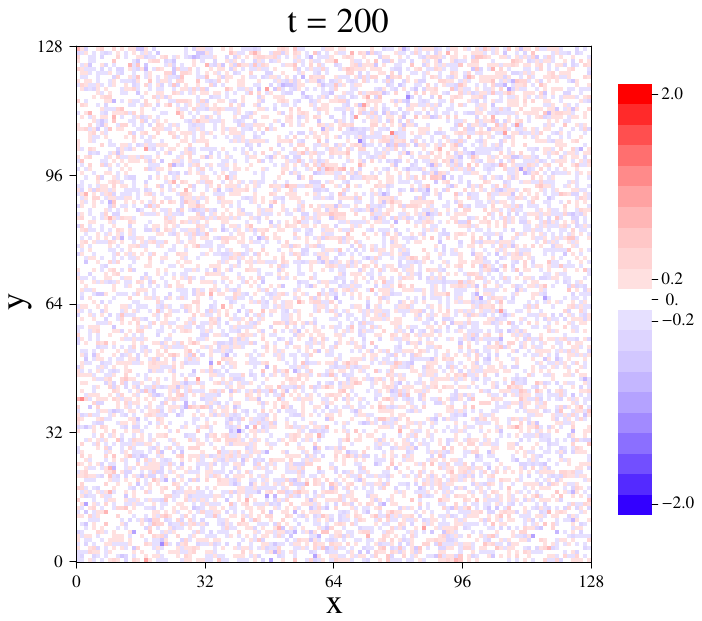}
\includegraphics[width=0.5\textwidth,bb=95 3 300 185]{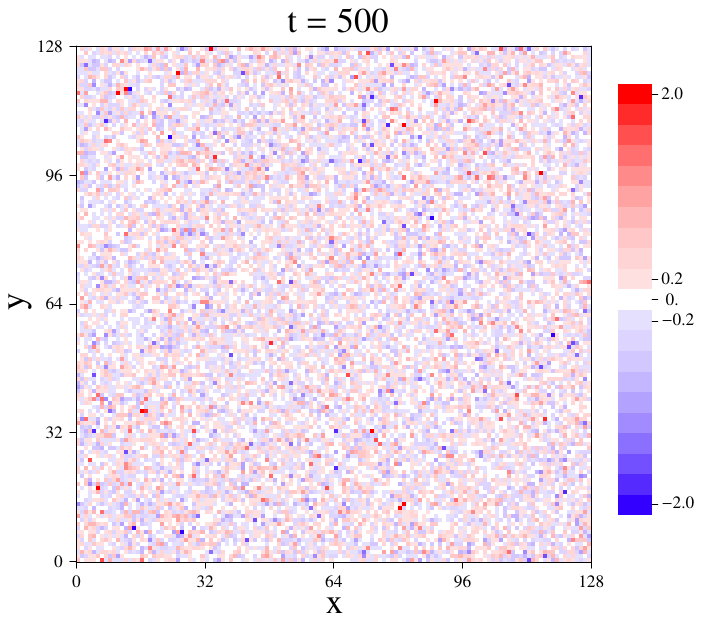}
\caption{\label{d4cpplnumsl}
Lepton number density as a function of the 2D lattice for time slices at $t=40,60,80,100,200,500$, for $\Delta\phi=4m$, $CP+$, initial flaton phase $\theta_\phi = \pi/4$.
}
\end{figure}

\clearpage

\begin{figure}
\includegraphics[width=0.5\textwidth,bb=95 3 300 185]{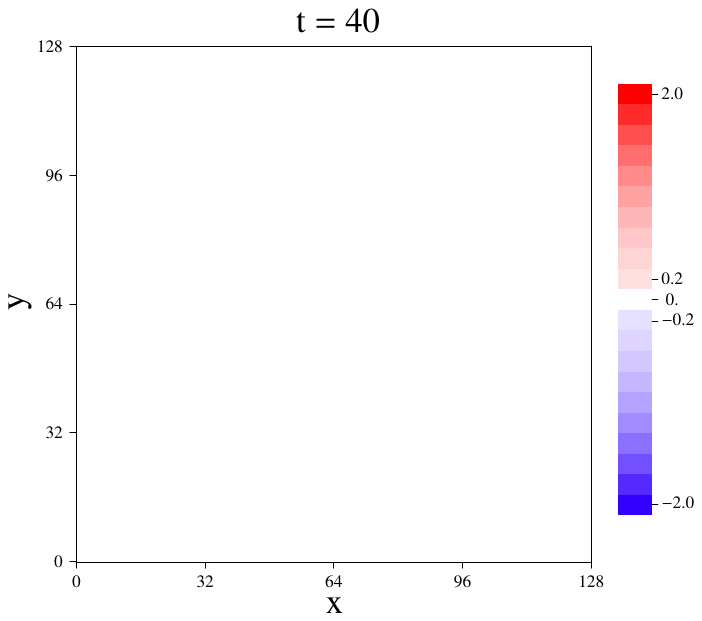}
\includegraphics[width=0.5\textwidth,bb=95 3 300 185]{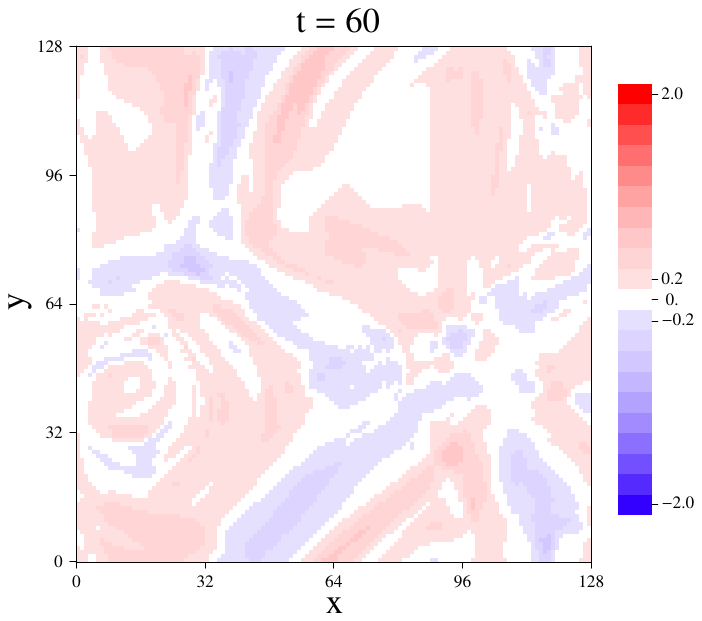}
\includegraphics[width=0.5\textwidth,bb=95 3 300 185]{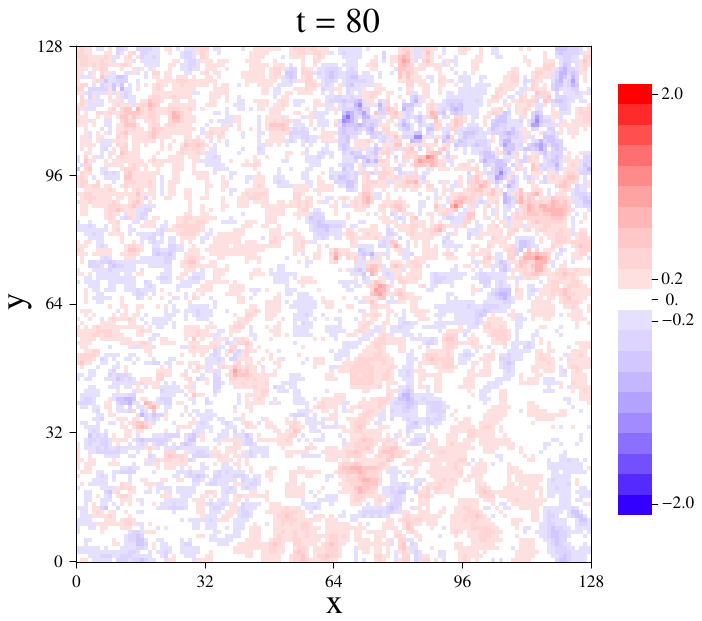}
\includegraphics[width=0.5\textwidth,bb=95 3 300 185]{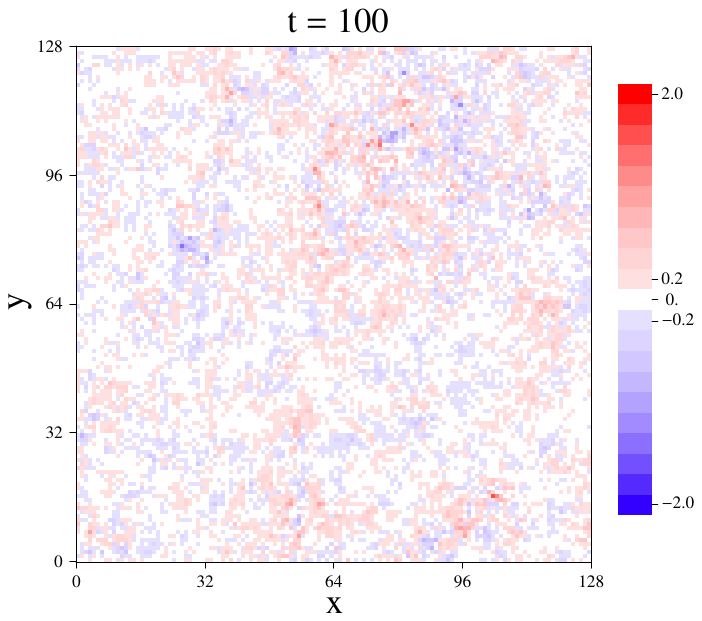}
\includegraphics[width=0.5\textwidth,bb=95 3 300 185]{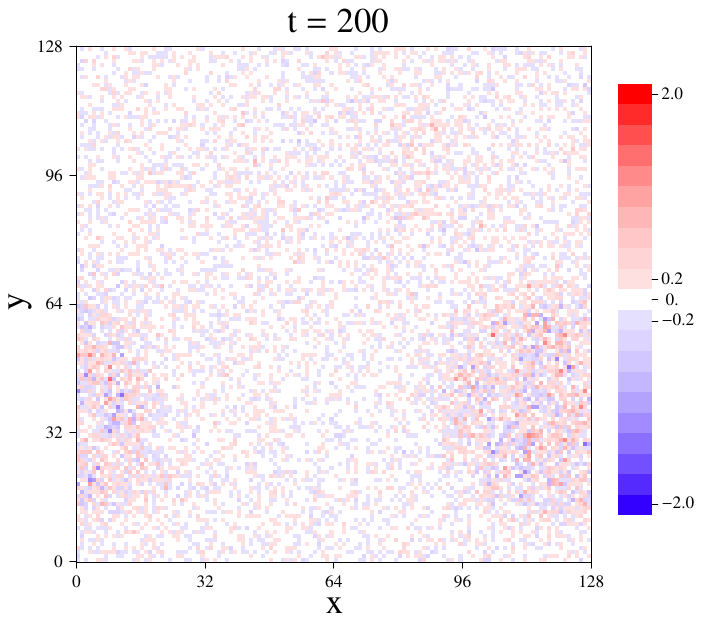}
\includegraphics[width=0.5\textwidth,bb=95 3 300 185]{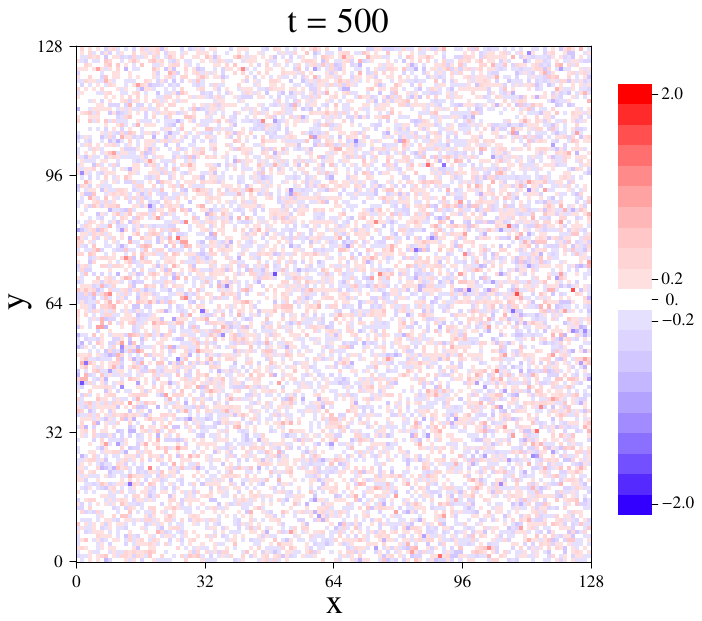}
\caption{\label{d0cpplnumsl}
Lepton number density as a function of the 2D lattice for time slices at $t=40,60,80,100,200,500$, for $\Delta\phi=0$, $CP+$.
}
\end{figure}

\clearpage


\begin{figure}
\includegraphics[width=0.5\textwidth,bb=110 3 310 146]{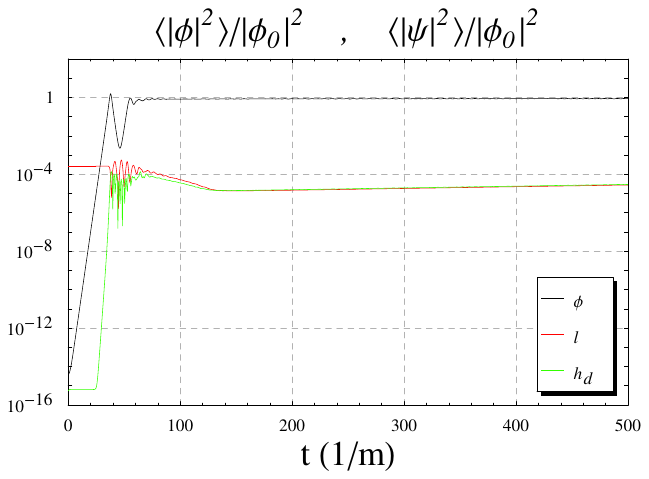}
\includegraphics[width=0.5\textwidth,bb=110 3 310 146]{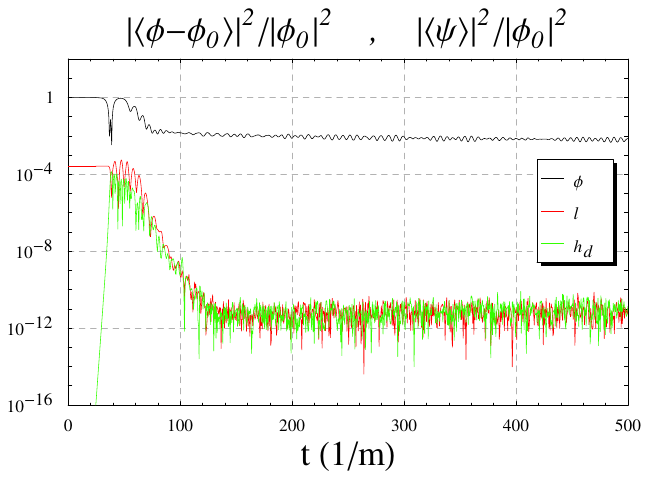}
\includegraphics[width=0.5\textwidth,bb=110 3 310 146]{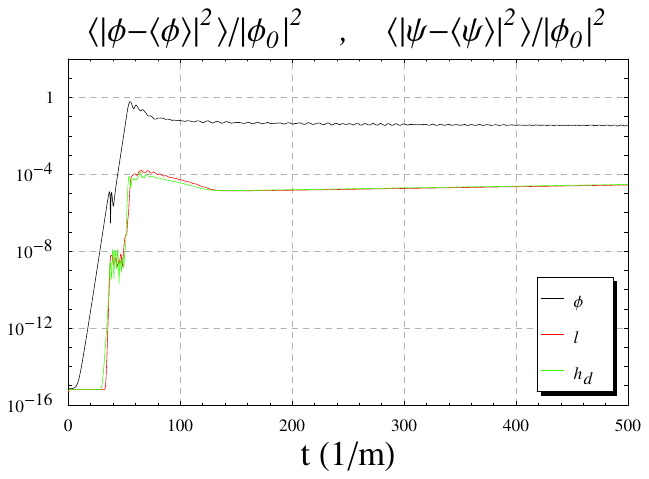}
\includegraphics[width=0.5\textwidth,bb=110 3 310 146]{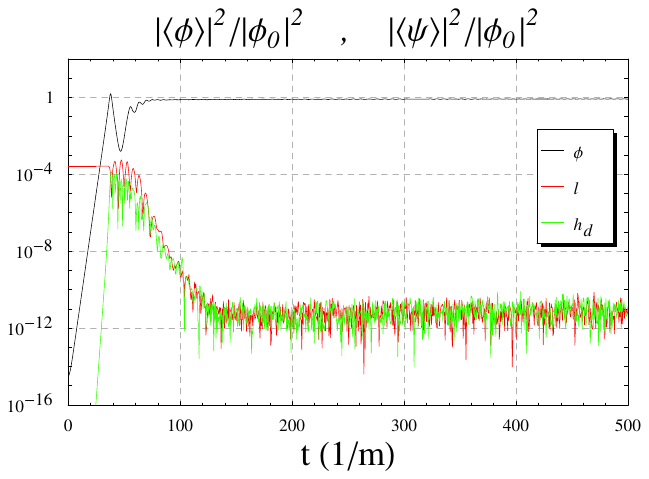}
\includegraphics[width=0.5\textwidth,bb=110 3 310 146]{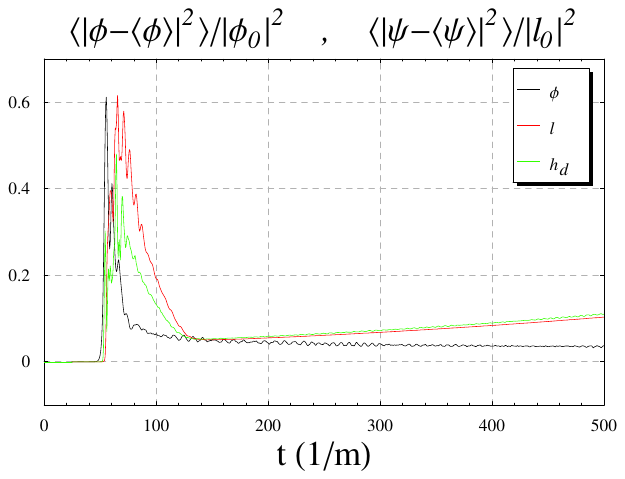}
\includegraphics[width=0.5\textwidth,bb=110 3 310 146]{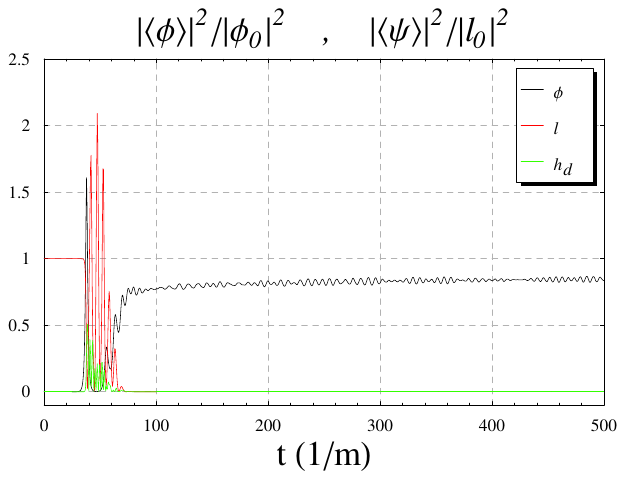}
\caption{\label{d4v}
Dispersions and variances averaged over the lattice, as a function of time, for $\Delta\phi=4m$, $\theta_\phi = \pi / 4$, $CP+$.
Left top - field amplitudes in log scale;
Left middle and bottom - dispersions in log and linear scales;
Right top - mean square with respect to vev in log scale;
Right middle and bottom - mean squares with respect to origin in log and linear scales.
}
\end{figure}

\clearpage

\begin{figure}
\includegraphics[width=0.5\textwidth,bb=110 3 310 146]{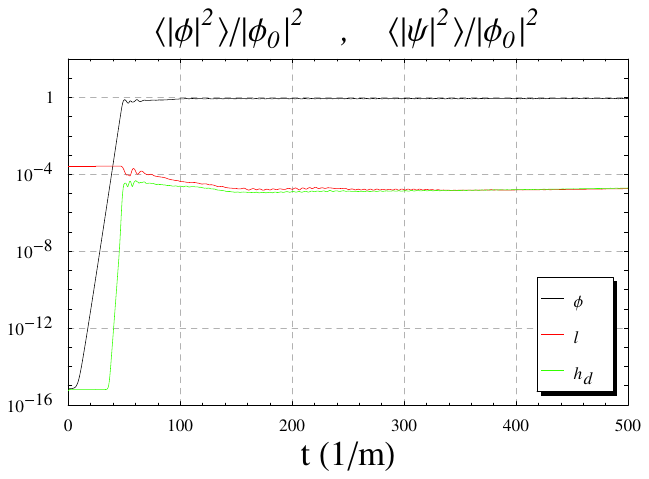}
\includegraphics[width=0.5\textwidth,bb=110 3 310 146]{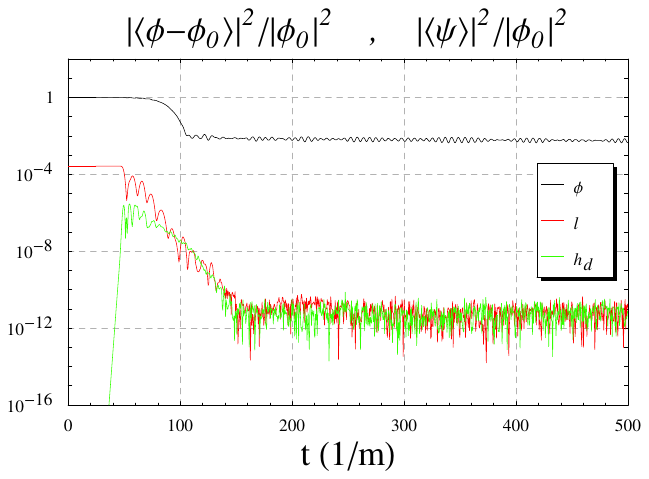}
\includegraphics[width=0.5\textwidth,bb=110 3 310 146]{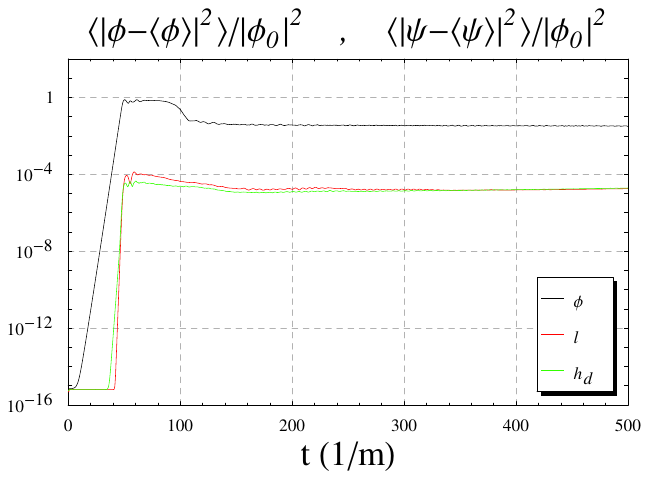}
\includegraphics[width=0.5\textwidth,bb=110 3 310 146]{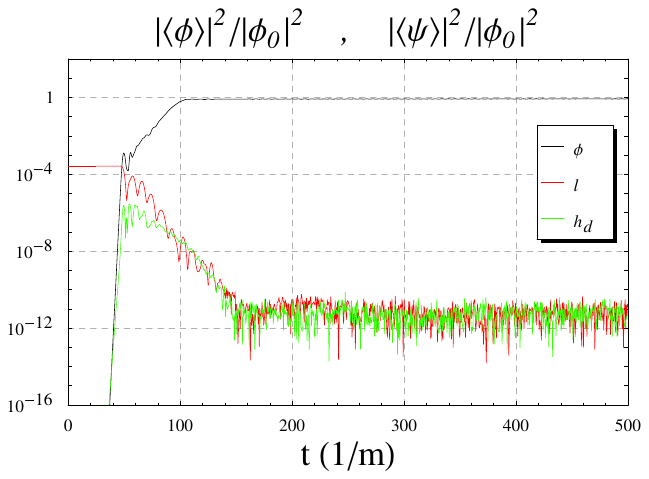}
\includegraphics[width=0.5\textwidth,bb=110 3 310 146]{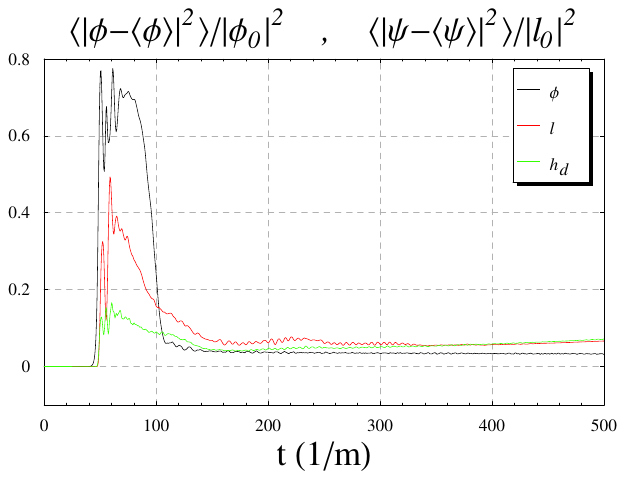}
\includegraphics[width=0.5\textwidth,bb=110 3 310 146]{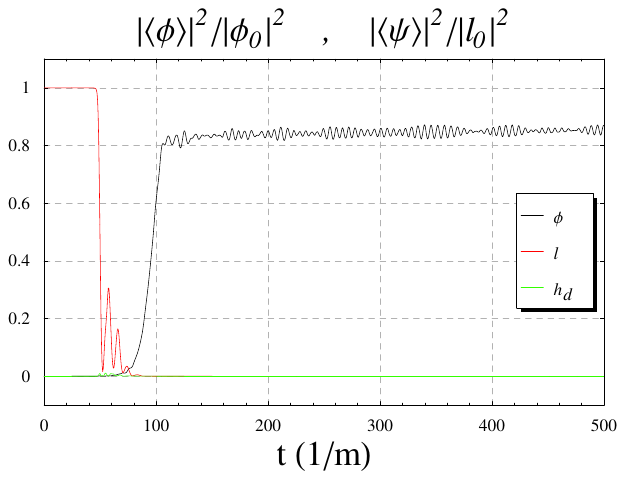}
\caption{\label{d0v}
Dispersions and variances averaged over the lattice, as a function of time, for $\Delta\phi=0$, $CP+$.
Left top - field amplitudes in log scale;
Left middle and bottom - dispersions in log and linear scales;
Right top - mean square with respect to vev in log scale;
Right middle and bottom - mean squares with respect to origin in log and linear scales.
}
\end{figure}

\clearpage


\begin{figure}
\includegraphics[width=0.5\textwidth,bb=95 3 305 192]{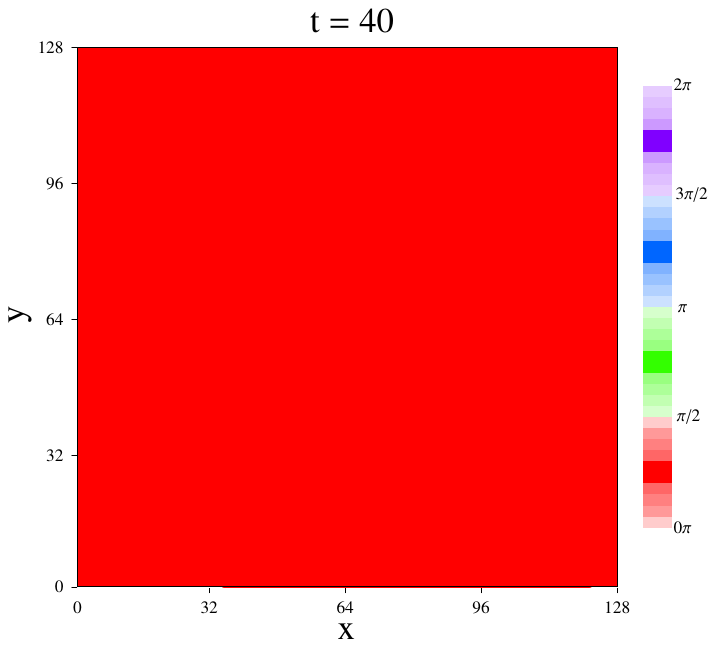}
\includegraphics[width=0.5\textwidth,bb=95 3 305 192]{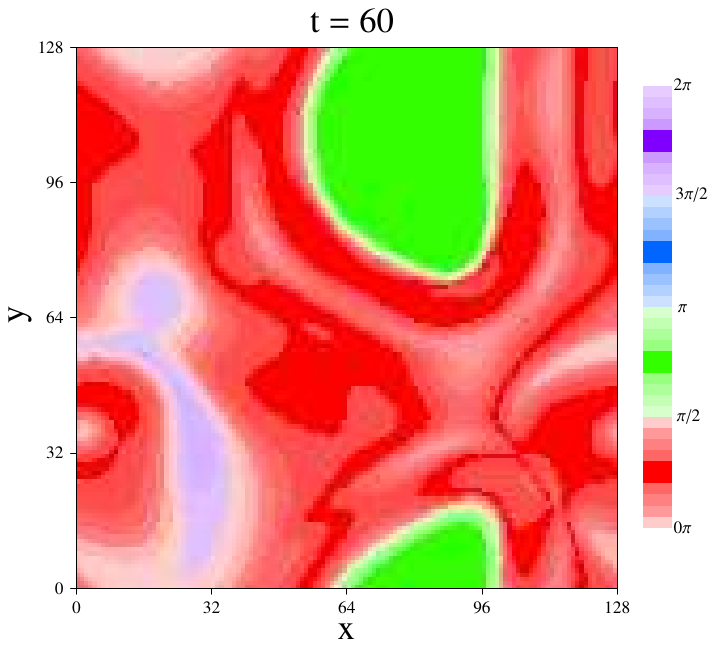}
\includegraphics[width=0.5\textwidth,bb=95 3 305 192]{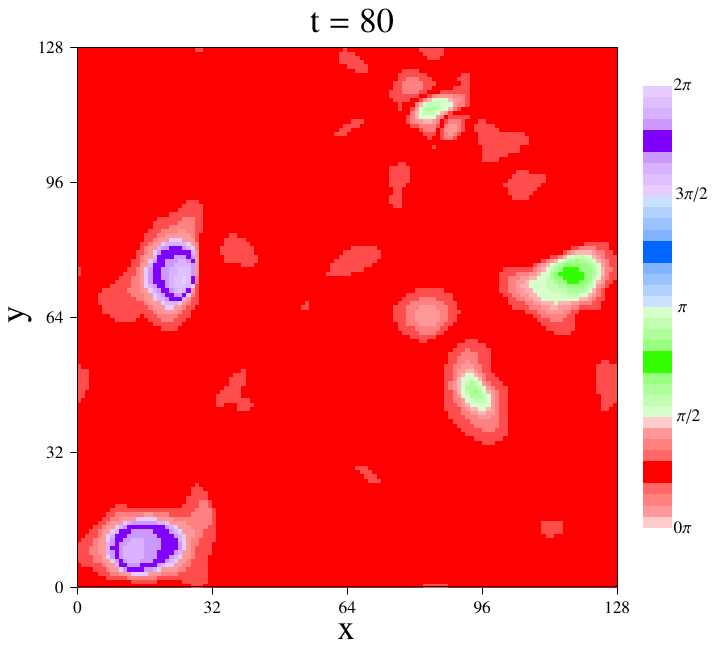}
\includegraphics[width=0.5\textwidth,bb=95 3 305 192]{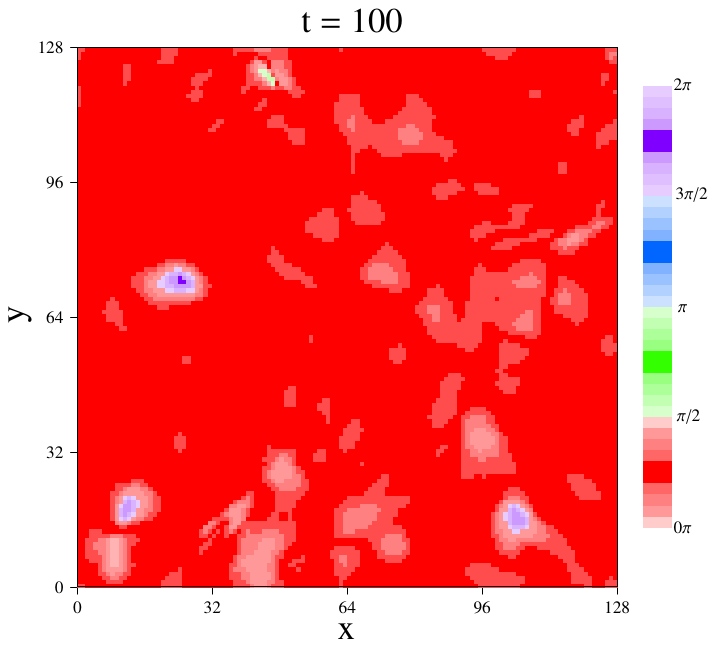}
\includegraphics[width=0.5\textwidth,bb=95 3 305 192]{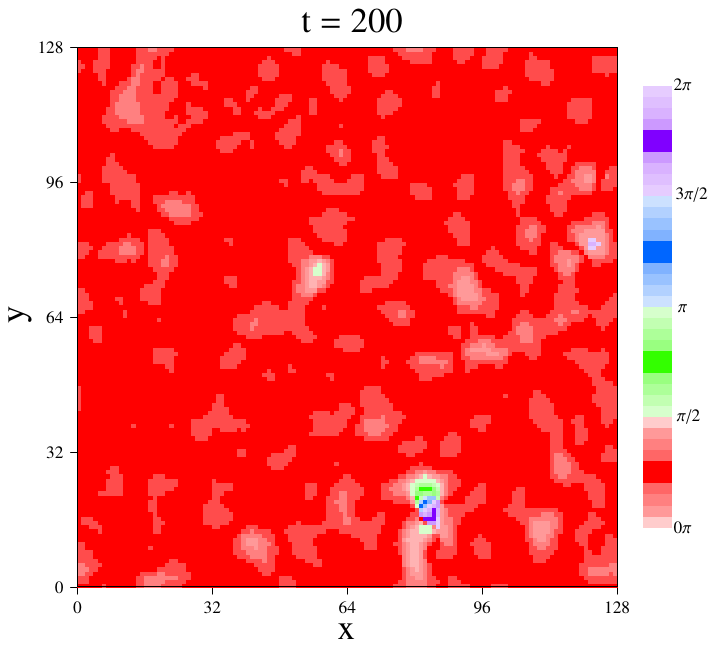}
\includegraphics[width=0.5\textwidth,bb=95 3 305 192]{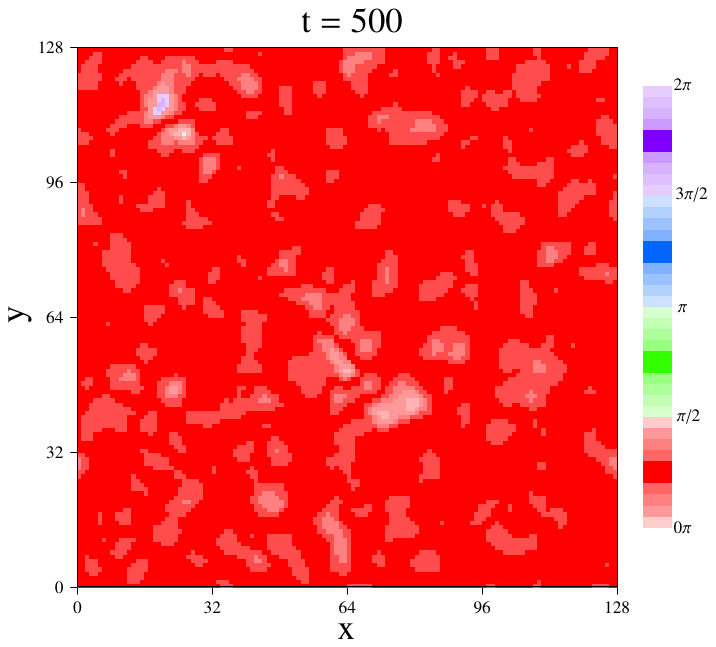}
\caption{\label{d4cppphasesl}
Flaton phase as a function of the 2D lattice for time slices at $t=40,60,80,100,200,500$, for $\Delta\phi=4m$, $\theta_\phi = \pi/4$, $CP+$.
}
\end{figure}

\clearpage

\begin{figure}
\includegraphics[width=0.5\textwidth,bb=95 3 305 192]{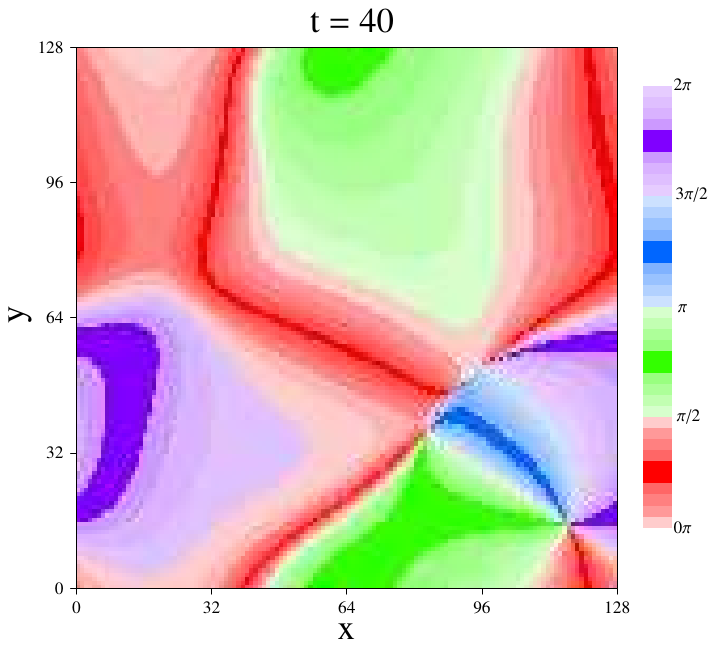}
\includegraphics[width=0.5\textwidth,bb=95 3 305 192]{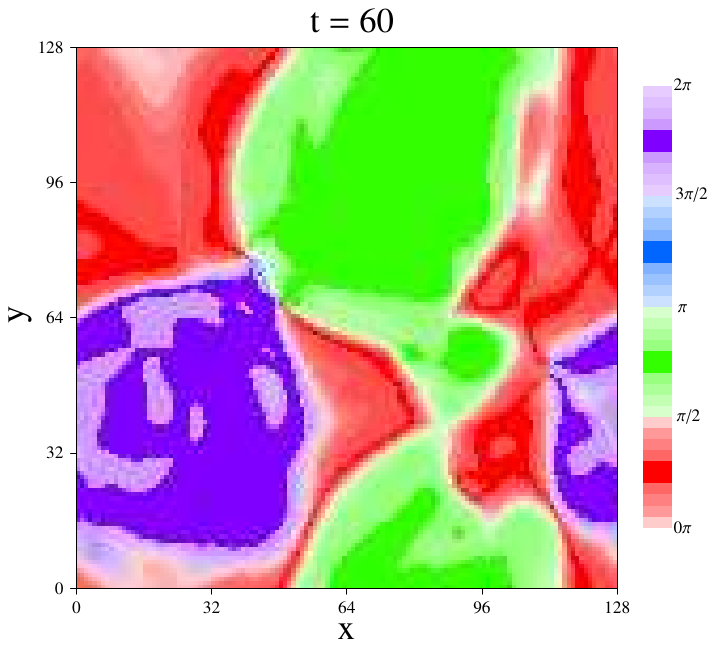}
\includegraphics[width=0.5\textwidth,bb=95 3 305 192]{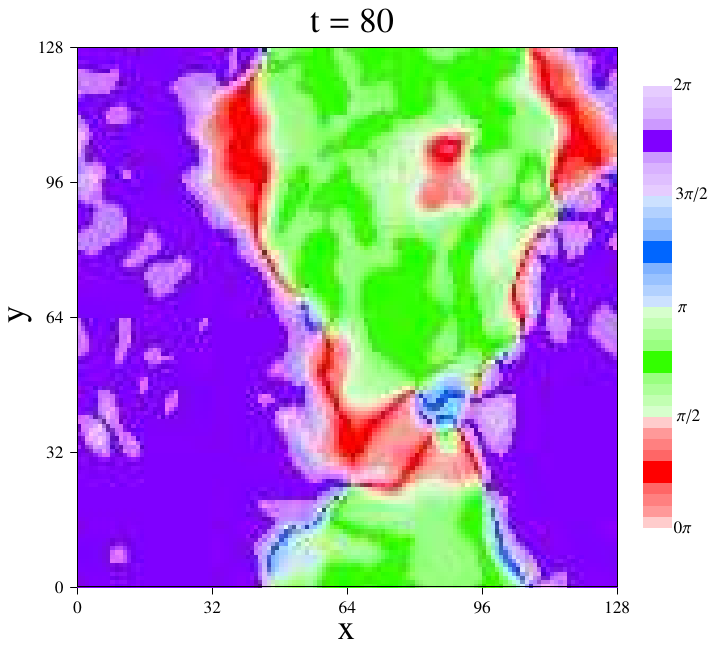}
\includegraphics[width=0.5\textwidth,bb=95 3 305 192]{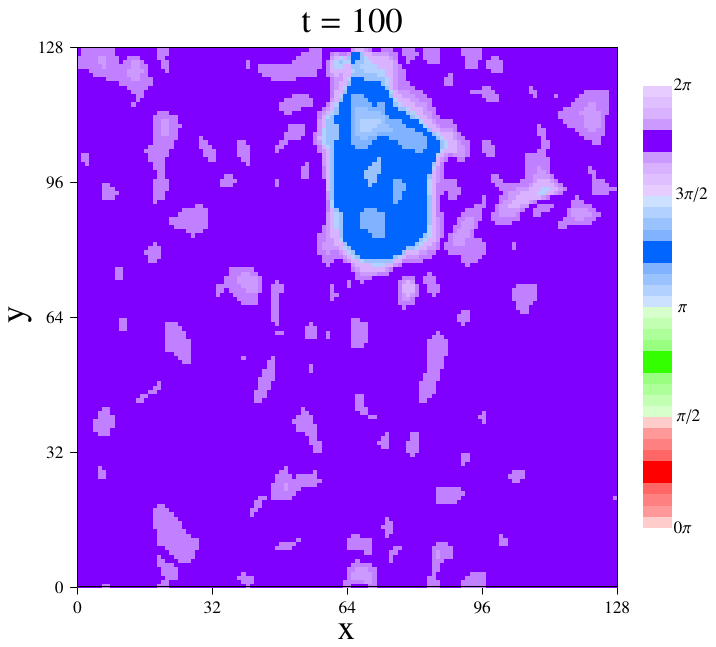}
\includegraphics[width=0.5\textwidth,bb=95 3 305 192]{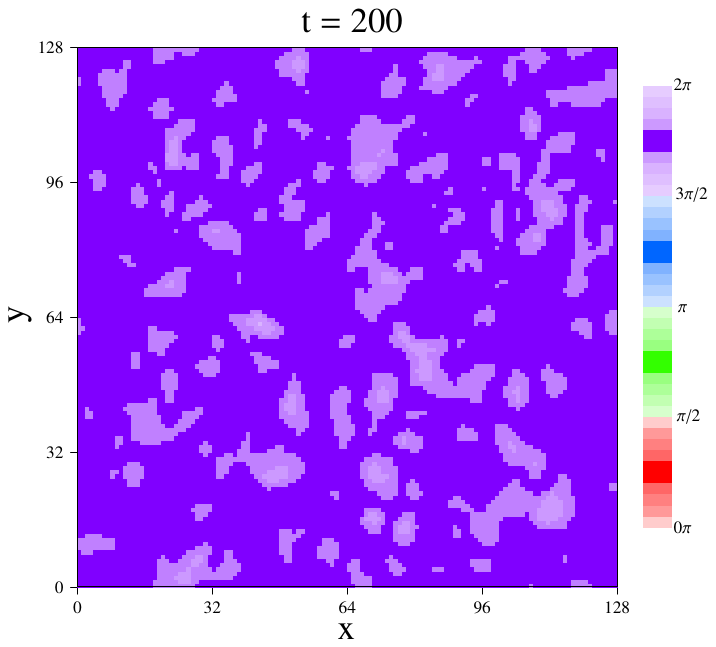}
\includegraphics[width=0.5\textwidth,bb=95 3 305 192]{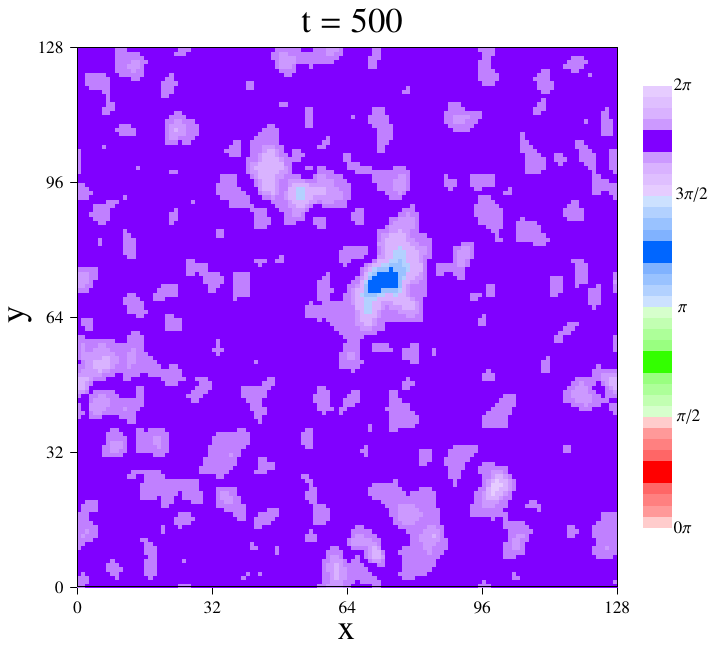}
\caption{\label{d0cppphasesl}
Flaton phase as a function of the 2D lattice for time slices at $t=40,60,80,100,200,500$, for $\Delta\phi=0$, $CP+$.
}
\end{figure}

\clearpage


\begin{figure}
\includegraphics[width=0.5\textwidth,bb=90 3 320 146]{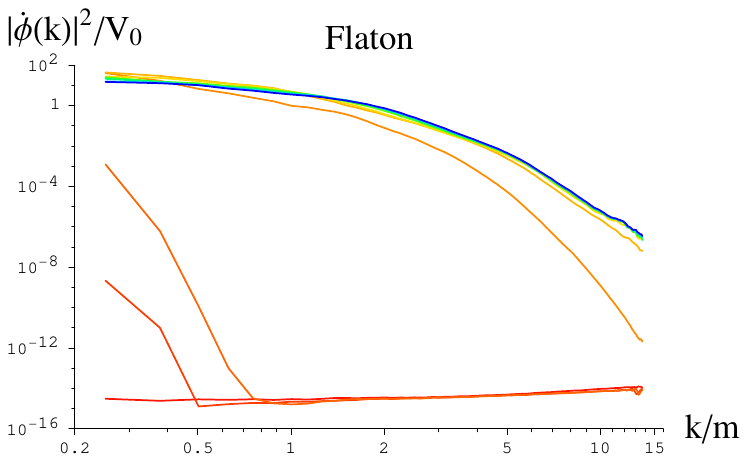}
\includegraphics[width=0.5\textwidth,bb=90 3 320 146]{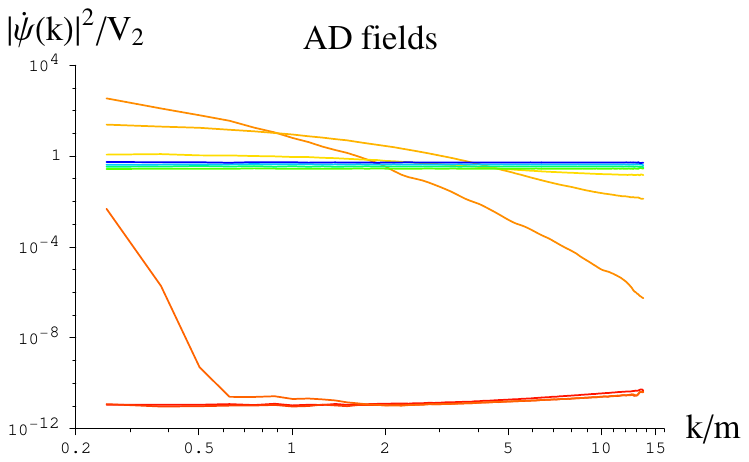}
\includegraphics[width=0.5\textwidth,bb=90 3 320 146]{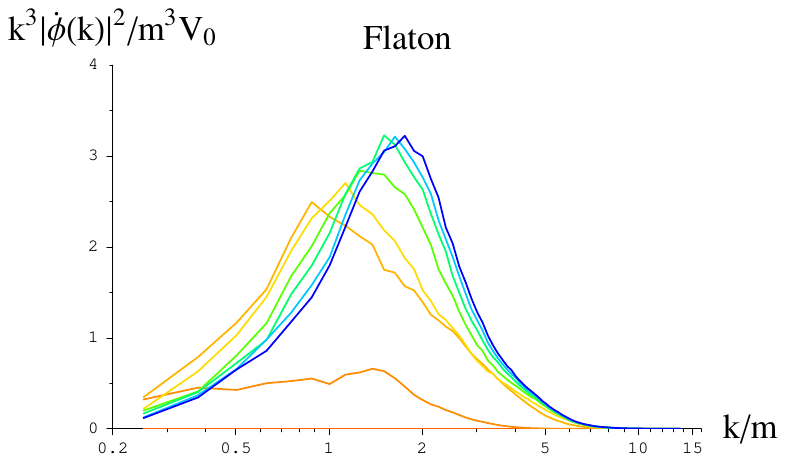}
\includegraphics[width=0.5\textwidth,bb=90 3 320 146]{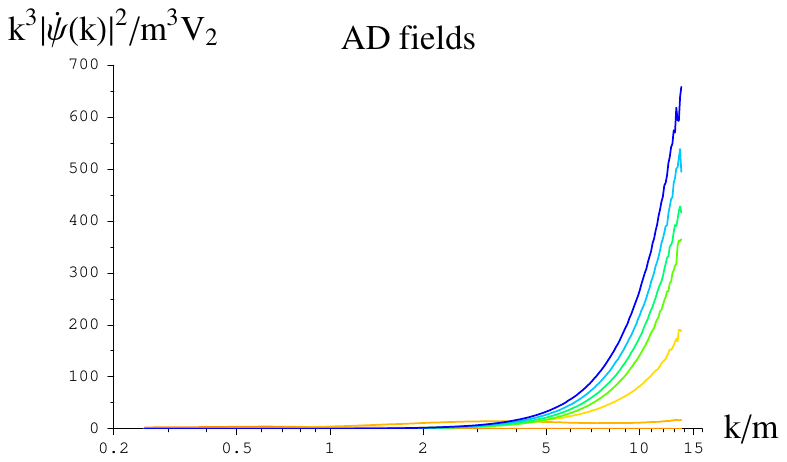}
\includegraphics[width=0.5\textwidth,bb=90 3 320 146]{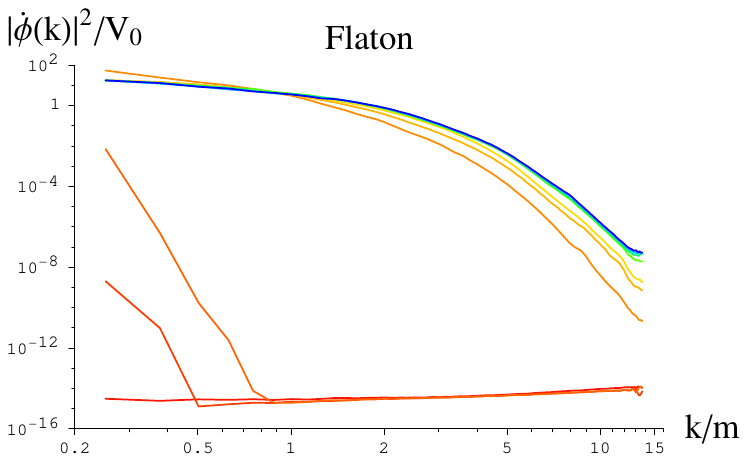}
\includegraphics[width=0.5\textwidth,bb=90 3 320 146]{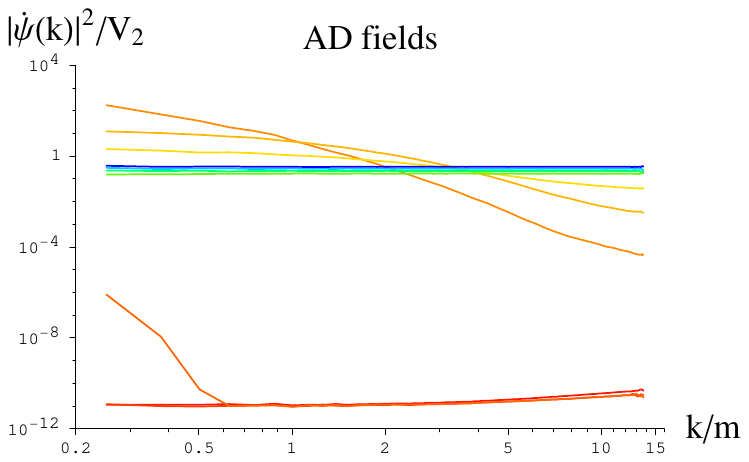}
\includegraphics[width=0.5\textwidth,bb=90 3 320 146]{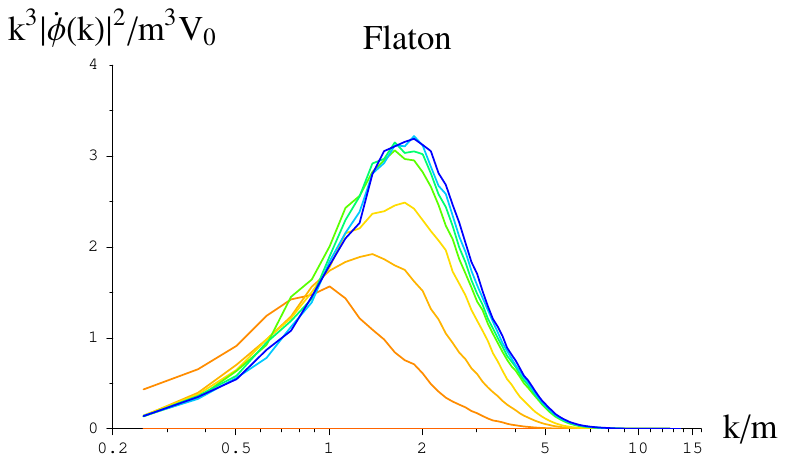}
\includegraphics[width=0.5\textwidth,bb=90 3 320 146]{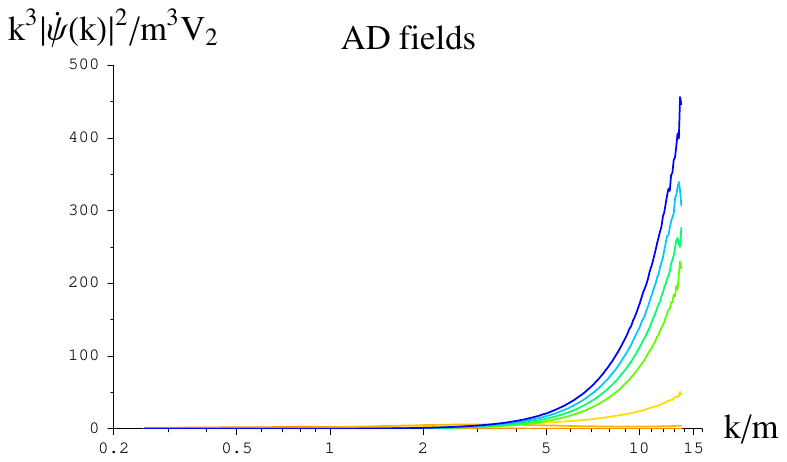}
\caption{\label{kspec}
Kinetic energy spectra of flaton and AD fields, at $t=0,20,40,60,80,100,200,300,400,500$ for $CP+$.
Thermally normalized so that a thermal spectrum is flat and energy normalized so that the area under the graph is the total energy from top to bottom.
Upper half: $\Delta\phi=4m$, $\theta_\phi = \pi/4$.
Lower half: $\Delta\phi=0$.
}
\end{figure}

\clearpage


\begin{figure}
\includegraphics[width=\textwidth,bb=90 3 325 146]{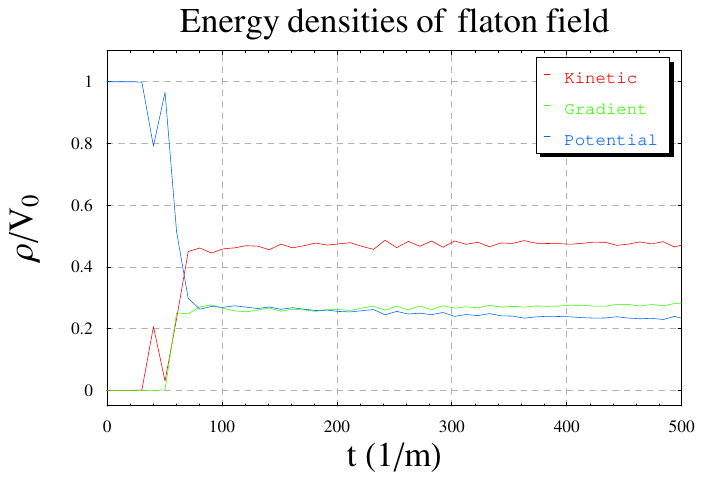}
\includegraphics[width=\textwidth,bb=90 3 325 146]{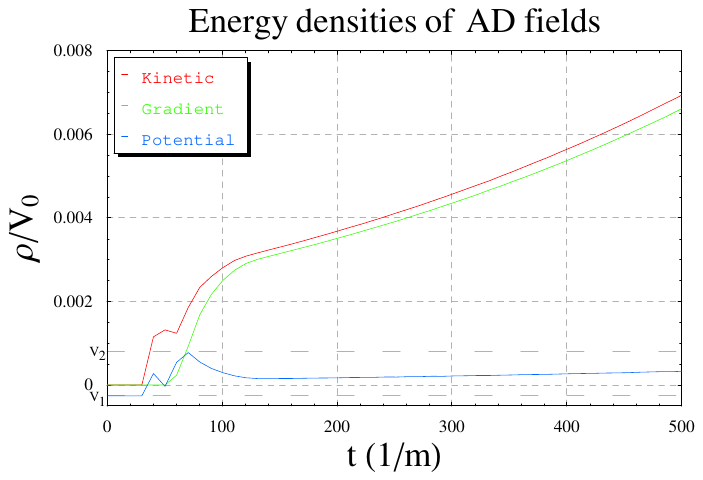}
\caption{\label{d4cppenergy}
Energy densities of the flaton as a function of time, for $\Delta\phi=4m$, $CP+$, $\theta_\phi = \pi/4$. From top to bottom at $t=500$: kinetic, gradient, potential.
}
\end{figure}

\clearpage

\begin{figure}
\includegraphics[width=\textwidth,bb=90 3 325 146]{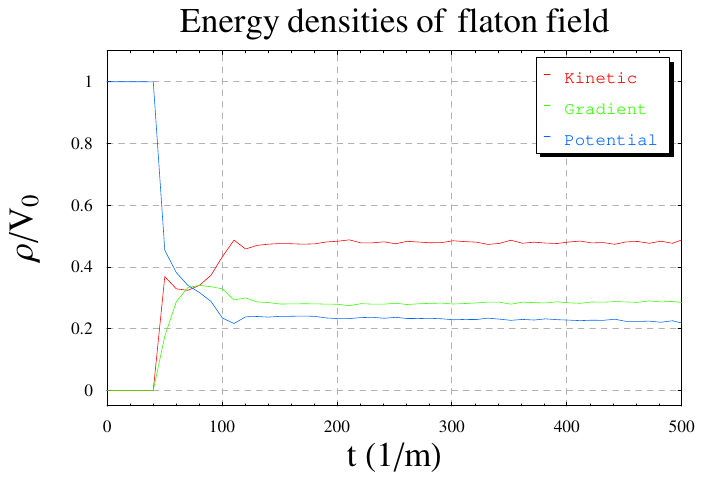}
\includegraphics[width=\textwidth,bb=90 3 325 146]{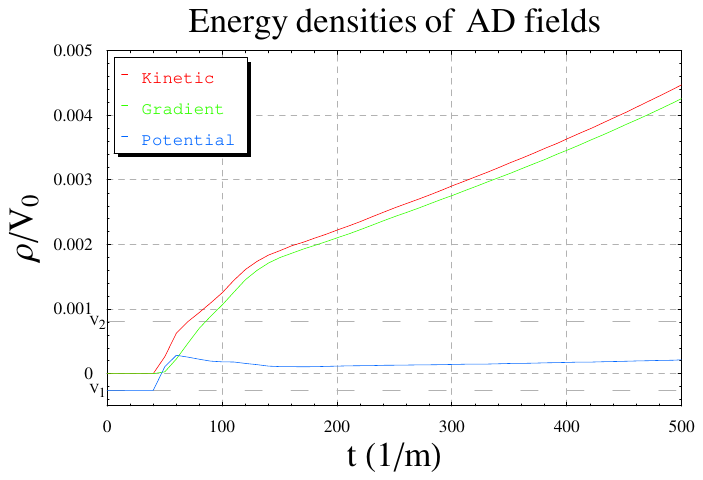}
\caption{\label{d0cppenergy}
Energy densities of the flaton and  AD fields as a function of time, for $\Delta\phi=0$, $CP+$. From top to bottom at $t=500$: kinetic, gradient, potential.
}
\end{figure}

\clearpage


\begin{figure}
\includegraphics[width=\textwidth,bb=90 3 325 140]{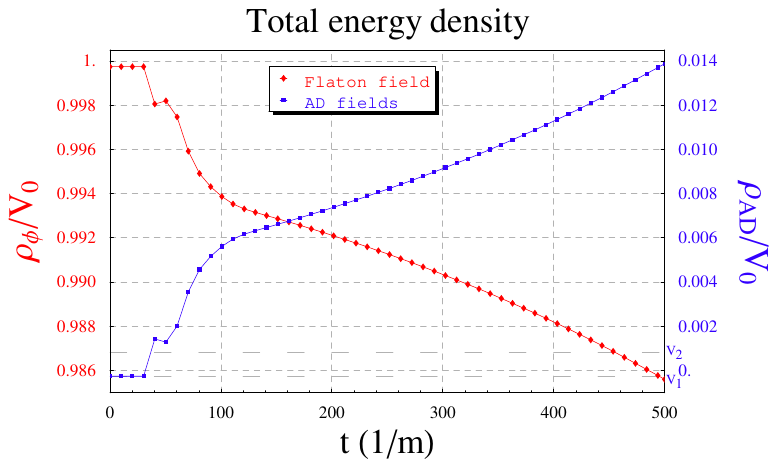}
\caption{\label{d4cppenergytot}
Energy flow between flaton and AD fields, for $\Delta\phi=4m$, $CP+$, $\theta_\phi = \pi/4$.
}
\end{figure}

\begin{figure}
\includegraphics[width=\textwidth,bb=90 3 325 140]{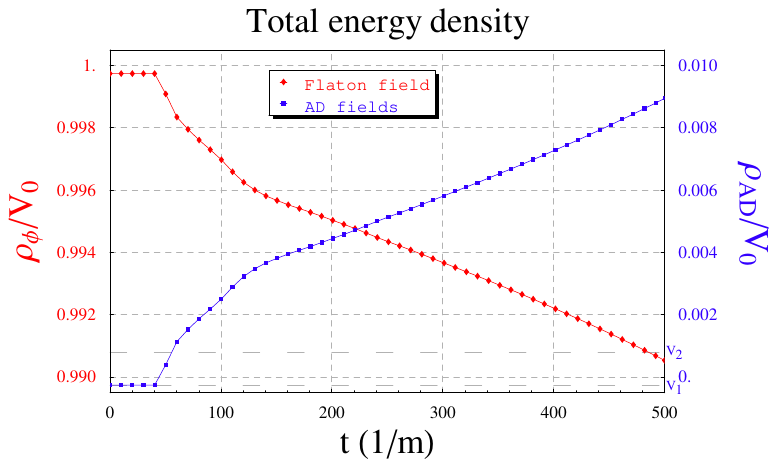}
\caption{\label{d0cppenergytot}
Energy flow between flaton and AD fields, for $\Delta\phi=0$, $CP+$.
}
\end{figure}

\clearpage


\begin{figure}
\includegraphics[width=0.5\textwidth,bb=90 3 320 146]{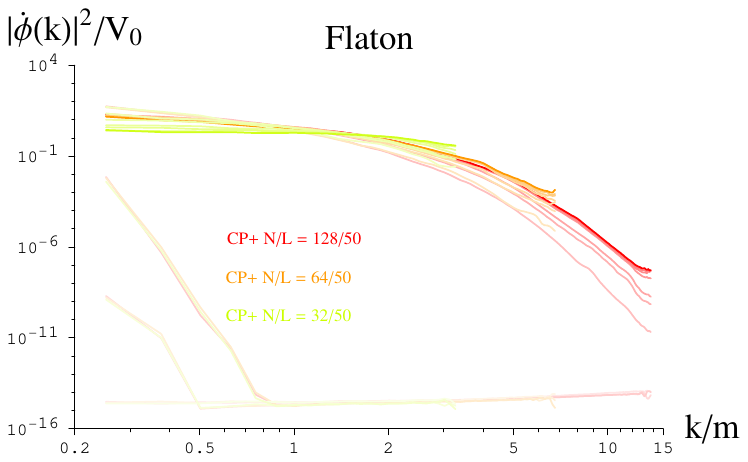}
\includegraphics[width=0.5\textwidth,bb=90 3 320 146]{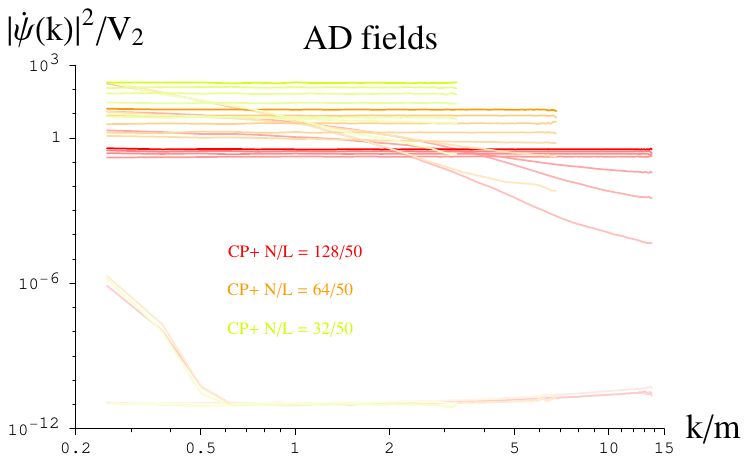}
\includegraphics[width=0.5\textwidth,bb=90 3 320 146]{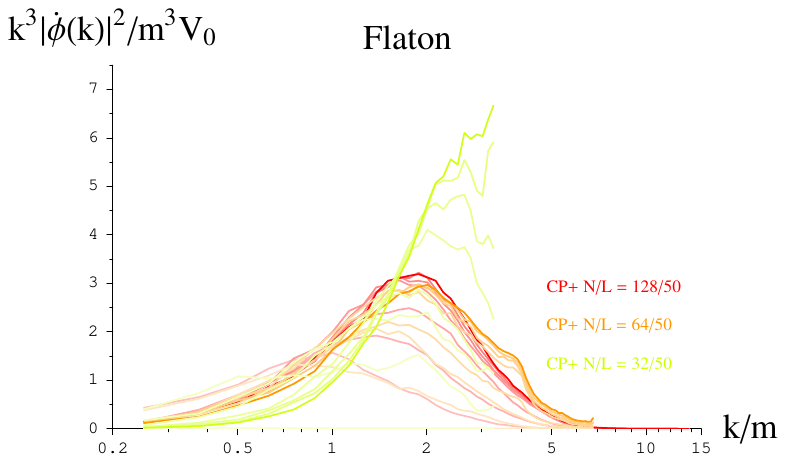}
\includegraphics[width=0.5\textwidth,bb=90 3 320 146]{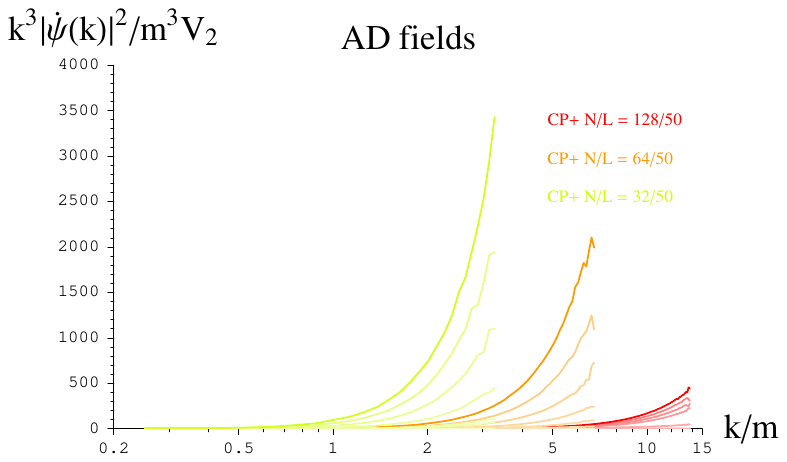}
\includegraphics[width=0.5\textwidth,bb=100 3 310 146]{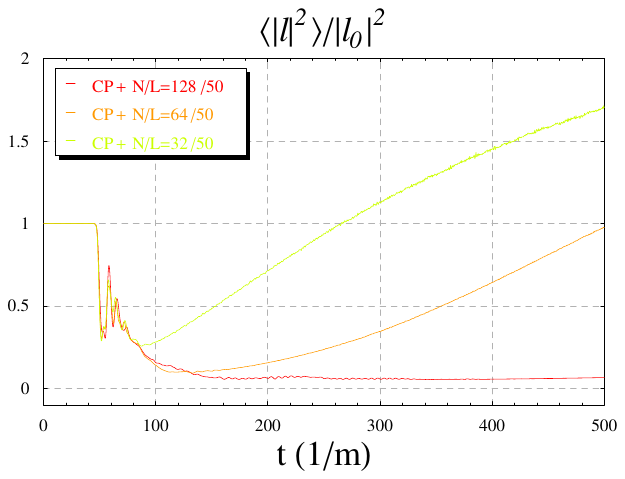}
\includegraphics[width=0.5\textwidth,bb=100 3 310 146]{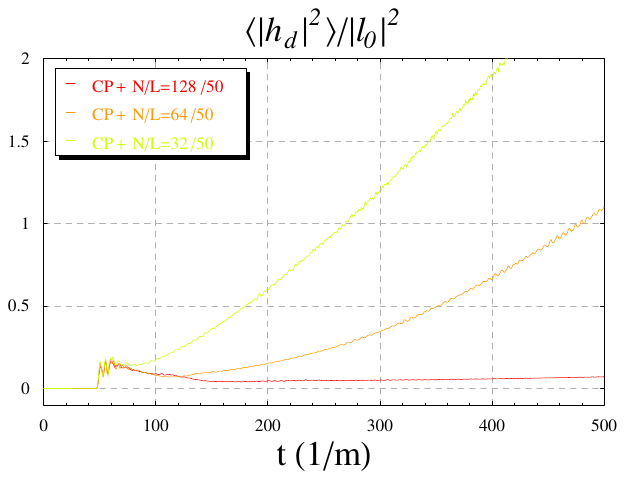}
\includegraphics[width=0.5\textwidth,bb=98 3 312 146]{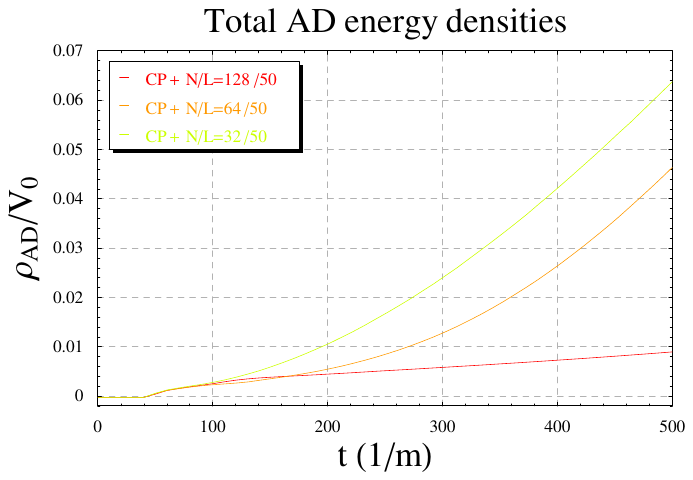}
\includegraphics[width=0.5\textwidth,bb=100 3 315 140]{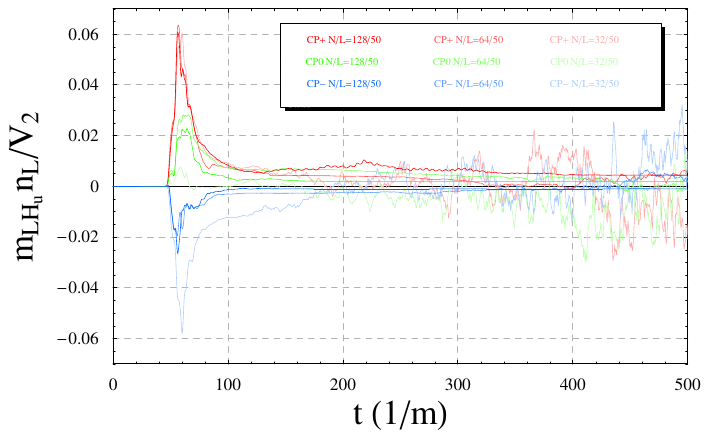}
\caption{\label{d0cppvslattice}
Dependence on lattice with $L = 50$ and $N = 32, 64, 128$ for $\Delta \phi = 0$, $CP+$.
Upper half: thermally normalized and energy normalized spectra in order at $t = 0,20,40,60,80,100,200,300,400,500$.
Lower half: variances, total energies and lepton numbers densities , averaged over the lattice, as a function of time.
The graph of the lepton number densities is including the cases of $CP0$ and $CP-$ as well.}
\end{figure}

\end{document}